\documentclass[aps,prc,preprint,groupedaddress,onecolumn,floatfix]{revtex4}
\pdfmapfile{-mpfonts.map}
\usepackage{amsmath}
\usepackage{amssymb}
\usepackage{amsfonts}
\usepackage{amscd}
\usepackage{bm}
\usepackage{bbold}
\usepackage{inter}
\usepackage{graphicx}
\usepackage{color}
\usepackage{empheq}

\def\beq{\begin{equation}}
\def\eeq{\end{equation}}
\begin{document}

\title{Path integrals, complex probabilities and the discrete Weyl representation} %Title of paper

\author{W.~N.~Polyzou}
\email{polyzou@uiowa.edu}
\thanks{This work supported by the U.S. Department of Energy,
  Office of Science, Grant \#DE-SC16457}
\affiliation{Department of Physics and Astronomy\\
  The University of Iowa\\
  Iowa City, IA 52242, USA}

\begin{abstract}

A discrete formulation of the real-time path integral as the
expectation value of a functional of paths with respect to a complex
probability on a sample space of discrete valued paths is explored.
The formulation in terms of complex probabilities is motivated by a
recent reinterpretation of the real-time path integral as the
expectation value of a potential functional with respect to a complex
probability distribution on cylinder sets of paths.  The discrete
formulation in this work is based on a discrete version of the Weyl
algebra that can be applied to any observable with a finite number of
outcomes.  The origin of the complex probability in this work is the
completeness relation.  In the discrete formulation the complex
probability exactly factors into products of conditional probabilities
and exact unitarity is maintained at each level of approximation.  The
approximation of infinite dimensional quantum systems by discrete
systems is discussed.  The method is illustrated by applying it to
scattering theory and quantum field theory.  The implications of these
applications for quantum computing is discussed.

\end{abstract}

\maketitle 

\bigskip

\section{Introduction}
\label{sec:intro}

Quantum computing has become a topic of current research because of
its potential for solving problems that are not accessible using
digital computers.  Quantum computations involve preparing an initial
state, evolving it in real time, and performing a measurement at a
later time; repeating this for a statistically significant number of
measurements.  An important step in this process is evolving the
prepared initial state using real-time evolution. Real-time path
integrals represent unitary time evolution as an integral over a
functional of classical paths.

Path integrals are formally derived using the Trotter product formula
\cite{Simon}, which expresses real-time evolution as the limit of a
product of time evolution over small time steps, making approximations
that preserve unitarity and become exact in the limit of small time
steps.  These are the same steps that quantum computing algorithms are
designed to replicate.  Real quantum computers are discrete quantum
systems with a finite number of qubits.  Path integral treatments of
real-time evolution in discrete systems can be interpreted as models
of quantum computers.

Feynman's interpretation of time evolution as an integral over paths
follows by inserting integrals over complete sets of intermediate
states at each time step and performing integrals over the
intermediate momentum variables.  What remains is the limit of
products of finite-dimensional Fresnel integrals that are interpreted
as integrals of a functional of classical paths.  This interpretation
has great intuitive appeal and is useful for generating perturbative
expansions and other results.  There are well-known difficulties with
the interpretation of time evolution as an integral over paths.  The
integrand involves oscillations that are not absolutely integrable,
and the volume element involves an infinite product of complex
quantities.  In addition, the ``action'' functional that appears in the
integrand has finite difference approximations of derivatives of the
paths, where $\Delta x$ does not get small as $\Delta t \to 0$.  The
problem with the integral interpretation of the path integral is that
there is no positive countably additive measure on the space on paths.
This is only a problem with interpretation since the Trotter limit
exists independent of the absence of an interpretation as an integral
over paths.

An alternative interpretation
\cite{Muldowney}\cite{Katya_1}\cite{Katya_2} was recently introduced
that replaces the integral over paths by the expectation value of a
functional of paths with respect to a complex probability on cylinder
sets of paths.  The concept of complex probabilities is neither
intuitive or familiar, however in the Euclidean (imaginary time) case
there is a path measure which can alternatively be interpreted as a
probability measure.  When imaginary-time evolution is replaced by
real-time evolution there is no countably additive positive measure on
the space of paths and the probabilities become complex.  The result
of \cite{Muldowney}\cite{Katya_1}\cite{Katya_2} is that by replacing
Lebesgue integration over intermediate states by Henstock
\cite{Henstock}\cite{bartle} integration, which can be used to treat
products of the Fresnel integrals, it is possible to make mathematical
sense of the complex probability interpretation, which results in a
global solution of the Schr\"odinger equation and a unitary
one-parameter time-evolution group.  This provides a rigorous
reinterpretation of real-time path integrals as an average over a
collection of classical paths, as suggested by Feynman.

In the real-time case the Gaussian integrals over momenta are replaced
by Fresnel integrals which are not absolutely integrable.  The
Henstock integral is an adaptive generalization of the Riemann
integral that performs the cancellation of oscillating quantities in
Fresnel integrals before adding them, resulting in finite integrals.
For Gaussian Fresnel integrals the Henstock integral gives the same
result that is obtained using contour integration.  When used in
real-time path integrals the integrals over the intermediate spatial
coordinates are approximated by the generalized Riemann sums, and
paths that have values in sequences of these intervals at different
time slices define the cylinder sets of paths.  In this interpretation
a complex probability is associated with free time evolution of the
paths in each cylinder set.  The resulting complex probability has
most of the properties of a real probability, except that it is not
positive and countable additivity is replaced finite additivity.

The purpose of this paper is to further explore applications based on
the complex probability interpretation of the path integral.  While
reference \cite{polyzou} was focused on quantum systems that act on
infinite dimensional Hilbert spaces, this work develops the complex
probability interpretation on finite ($M$) dimensional Hilbert spaces.
The finite dimensional case is more closely related to quantum
algorithms.  In the finite dimensional case the complex probabilities
are intuitive and the Fresnel integrals are replaced by finite sums.
Also in the finite dimensional case the complex probabilities exactly
factor into products of conditional probabilities at each time step.
Cylinder sets of paths are identified with ordered sequences of
discrete eigenvalues of quantum observables at different intermediate
times.  Most of the ``paths'' in a cylinder set are nowhere continuous.
An alternative quantum mechanical interpretation is to treat the sequence
of eigenvalues as labels for an ordered sequence of quantum mechanical
transition probability amplitudes that result from the dynamics
between each time step.    The complex
probabilities used in this application are due to the completeness
sums over intermediate states.  Systems with continuous degrees of
freedom, which are treated in \cite{polyzou}, can be approximated as
limits of discrete systems, as they would be in a quantum computer.

Below is a brief outline of the structure of this paper.  The ultimate
goal is to illustrate by example how discrete methods can in principle
be used to perform calculations of scattering observables and the
dynamics of quantum fields on a quantum computer.  The approach of
this work is bottom up in the sense that the continuum dynamics is
treated as the limiting case of a sequence of discrete models.  Real
time evolution of systems on finite dimensional Hilbert spaces is
naturally formulated using a complex probability representation.
In this work the dynamics is expressed as an average over a space
of paths, the formal path integral is replaced by the expectation
value of a functional of ``classical paths'' with respect to a complex
probability distribution on cylinder sets of paths.  This
reinterpretation of the path integral was developed in
\cite{Muldowney}\cite{Katya_1}\cite{Katya_2}.  In the continuum case
complex probability densities are related to products of free particle
transfer matrices.  While the Fresnel integrals that appear in these
transfer matrices are not Riemann integrable, they can be computed
using an adaptive generalization of the Riemann integral due to
Henstock \cite{Henstock}\cite{bartle}\cite{Muldowney}, which is used to define the complex probabilities.  The
advantage of the application to discrete systems is that the Fresnel
integrals are replaced by finite sums. Since the complex probabilities
play a prominent role in this work, their essential properties are
discussed briefly in section \ref{sec:complexp}.  The general treatment of
discrete systems is discussed in section \ref{sec:weyl} and \ref{app2}.  The starting point
is a general observable with a finite number of outcomes.  It is used,
following a method due to Schwinger, to construct an irreducible pair
of unitary operators; one commuting with the original observable and
one complementary operator with the property that any operator in the
Hilbert space can be expressed as a finite degree polynomial in both
operators.  The path integral representation of the time evolution
operator for these discrete systems as the expectation value of a
potential functional on cylinder sets of discrete-valued paths is
developed in section \ref{sec:findim} and \ref{app1}.  The discrete form has the advantage that
the complex probability associated with $N$ time steps exactly factors
into a product of conditional probabilities for single time steps.
This factorization is used to exactly replace the sum over a large
number of cylinder sets of paths by powers of a finite dimensional matrix.  While this
factorization is already useful, quantum computers are normally
formulated on a Hilbert space that is generated by tensor products of
qubits.  After a short discussion contrasting the difference between
classical and quantum computers in section \ref{sec:logic}, the formulation of the discrete path
integral in section \ref{sec:findim} is expressed in terms of qubits
in section \ref{sec:qubit}.  This is
achieved by starting with an observable on a $M=2^L$ dimensional
Hilbert space.  The irreducible pair of unitary operators is expressed
in terms of a collection of L pairs of complementary unitary operators
that act on individual qubits.  The elementary unitary operators can
be identified with standard quantum gates.

Most problems of interest for quantum computing, like
scattering and the dynamics of quantum fields, involve operators with a
continuous rather than discrete spectrum.  The approximation of a
quantum system on an infinite dimensional Hilbert space as
the limit of finite-dimensional systems discussed in section
\ref{cont}.  The resulting formulation of the real-time ``path
integral'' is given in section \ref{path}.  The application of the formalism to
a simple model of
a particle scattering off of a repulsive Gaussian potential is
discussed in the section \ref{scatt}.  The calculations utilize
time-dependent methods which involve strong limits.  Narrow wave
packets in momentum are used to extract sharp-momentum transition
matrix elements. In section \ref{qft} the application of
these methods to local field theories is discussed.
For field theories an additional discretization of the field amplitudes
is needed for computations.  In this
section a wavelet basis is used to represent the field in terms of
an infinite number of discrete modes.  The wavelet representation
is an exact representation of the field that 
replaces the fields, as operator valued distributions, by an infinite
collection of well-defined, almost local, canonically conjugate
pairs of operators.  In this representation singular products of fields
are replaced by infinite sums of products of well-defined operators.
The construction of the wavelet basis is discussed in \ref{app3}.
The wavelet
representation has natural volume and resolution truncations.
Computations require truncations to a finite number of degrees of
freedom.  The discrete path integral methods are used to compute the
time evolution of a $\phi^4(x)$ field with a Hamiltonian truncated to
two discrete modes.  While this is a drastic truncation, it illustrates how
these discrete methods can used to model the dynamics of fields.

\section{Complex probabilities}
\label{sec:complexp}
    
A complex probability system is defined
by a sample set $S$ and a complex valued function $P$ on subsets of
$S$ with the
properties
\beq
P(S) =1 .
\label{cp1}
\eeq
\beq
P(S_i)+P(S_i^c)= 1
\label{cp2}
\eeq
where $P(S_i)$ is the complex probability assigned
to the subset $S_i$ of $S$ 
and  $S_i^c$ is the complement of $S_i$ in $S$. 
For a finite set of disjoint subsets of $S$
\beq
S_i\cap S_j = \mbox{$\emptyset$ }, \qquad  i \not=j  \qquad P(\cup S_i) =\sum_i P(S_i).
\label{cp3}
\eeq
In the applications that follow the sample set, $S$, will be the set of
cylinder sets of ``paths'' that have discrete values at different times.
A cylinder set is the collection of ``paths'' that have
specific discrete values at finite collections of times between $0$
and $t$.  Since $P(S_i)$ is complex, equation (\ref{cp3}) cannot be
extended to countable non-intersecting subsets, which is where complex
probabilities differ from ordinary probabilities
\cite{Muldowney2}\cite{Muldowney}.

In this work each cylinder set is associated with a collection of
paths that have specific discrete values at each time step.  By
including additional time steps, the original set of paths is
decomposed into subsets of paths with different values at the new
intermediate times.  Summing over the complex probabilities for these
subsets of cylinder sets with different intermediate times
gives the complex probability associated with
the original cylinder set.  This property demonstrates that the
complex probability is defined consistently for any partition of the
interval $[0,t]$ into a finite collection of time steps.  The
treatment of complex probabilities for continuously infinite
dimensional path spaces is treated in \cite{Muldowney}.  The extension
of the notion of complex probabilities to paths with continuous values
based on the Henstock integral
\cite{Henstock}\cite{bartle}\cite{Gill}, was used in
\cite{Katya_1}\cite{Katya_2}.

\section{Schwinger's discrete Weyl algebra}
\label{sec:weyl}

This section summarizes Schwinger's \cite{schwinger} method for
constructing an irreducible algebra of complementary unitary operators
for quantum systems of a finite number of degrees of freedom.  This
construction generates a finite degree of freedom version of the Weyl
(exponential) form of the canonical commutations relations.  This
algebra can be used to build discrete models of any finite quantum system.
%This construction is essentially the same as the treatment of the
%quantum Fourier transform discussed in \cite{nielsen} and elsewhere.
This section summarizes the main results of this construction.  Details
of Schwinger's construction are presented in \ref{app2}.

The construction starts with a quantum observable $X$ with $M$
orthonormal eigenvectors, $\vert m \rangle$, associated with $M$
possible measurement outcomes, $x_m$.  $X$ is a $M \times M$ Hermitian
matrix on a $M$-dimensional complex Hilbert space ${\cal H}$.  The
outcomes are the eigenvalues of $X$:
\beq
X \vert m \rangle =x_m \vert m \rangle  \qquad m=1,\cdots, M .
\label{s1}
\eeq
The eigenvectors  can be chosen to form an orthonormal basis on ${\cal H}$.
Schwinger defines a unitary operator $U$ on ${\cal H}$ that cyclically
shifts the eigenvectors of $X$:
\beq
U\vert m \rangle = \vert {m+1} \rangle \qquad m<M
\qquad
U\vert M \rangle = \vert 1 \rangle .
\label{s2}
\eeq
The labels $m$ on the eigenvectors are treated
as integers mod $M$ so 0 is identified with $M$, $1$ with $M+1$ etc..
%$U$ defined by (\ref{s2}) is unitary since
%\beq
%UU^{\dagger}= \sum_{m=1}^M U\vert m \rangle \langle m \vert U^{\dagger}=
%\sum_{m=1}^{M-1} \vert {m+1} \rangle \langle {m+1} \vert 
%+ \vert {1} \rangle \langle {1} \vert =
%\sum_{m'=1}^M \vert {m'} \rangle \langle {m'} \vert = I .
%\label{s3}
%\eeq
Since $M$ applications of $U$ leaves all $M$
basis vectors, $\vert m \rangle$,  unchanged, it follows that $U^M=I$.
Since $U^k\vert m \rangle$ are independent for all $0<k \leq M$, there are no
lower-degree polynomials in $U$ that vanish, so 
$P(\lambda) = \lambda^M-1 =0$ is the characteristic polynomial of $U$.
The eigenvalues $\lambda$ 
of $U$ are the $M$ roots of $1$:
\beq
\lambda = u_m = e^{2\pi  m i\over M}
\label{s3}
\eeq
with orthonormal eigenvectors $\vert \bar{m} \rangle$:
\beq
U \vert \bar{m} \rangle = u_m \vert \bar{m} \rangle
%\qquad u_m = e^{\frac{2 \pi i m}{M}}
\label{s4}
\eeq
\beq
\langle \bar{m} \vert \bar{n} \rangle = \delta_{mn}.
\label{s5}
\eeq
Next Schwinger defines a second unitary operator that cyclically
shifts the eigenvectors of $U$ in the opposite direction
\beq
V \vert \bar{n} \rangle = \vert \bar{n-1} \rangle, \qquad n \not=1,
\qquad
V \vert \bar{1} \rangle = \vert \bar{M} \rangle . 
\label{s14}
\eeq
By definition 
\beq
V^M=I
\label{s15}
\eeq
and 
\beq
V \vert v_m \rangle = v_m \vert v_m \rangle
\qquad v_m = e^{2 \pi i m \over M}.
\label{s16}
\eeq
These definitions do not fix the phase of the eigenvectors of $U$ and $V$.
The phases can be chosen so
$\vert v_k \rangle = \vert k \rangle$ are the eigenvectors of the original
observable, $X$.

The operators $U$ and $V$ (see \ref{app2}) have the
following properties:

\begin{itemize}
\item[1.] Any operator commuting with both $U$ and $V$ is a constant multiple of the identity.

\item[2.] $UV= VUe^{-i \frac{2 \pi}{M}}$  

\item[3.] Any operator $O$ can be expressed as a degree $(M-1)\times(M-1)$ polynomial in $V$ and $U$
\[
O = \sum_{m,n=0}^{M-1}c_{mn} U^m V^n = \sum_{m,n=0}^{M-1}d_{mn} V^m U^n.
\]
  
\end{itemize}

\section{``Path integrals'' for finite dimensional systems}
\label{sec:findim}

This section considers a general system of $M<\infty $
degrees of freedom.  $X$ is the observable of interest.  It evolves
under the influence of a Hamiltonian, $H$, which is a Hermitian matrix
on the $M$-dimensional Hilbert space.  A measurement of $X$ gives one
of its eigenvalues, $x_n$.  The $M$ eigenvectors of $X$ are chosen to
be orthonormal.  The notation, $\vert n \rangle$ for $1 \leq n \leq M$ is
used to label these eigenvectors. $X$ will be called  a ``coordinate
operator'' in this section, but in principle it can be any Hermitian
operator on ${\cal H}$.

The quantity of interest is the probability amplitude for a transition
from an initial eigenstate $\vert n_i \rangle$ of $X$ to
a final eigenstate $\vert n_f \rangle$ after time $t$.  This
is given by the matrix element,
\beq
\langle n_f \vert e^{-iHt } \vert n_i \rangle ,
\label{f1}
\eeq
of the unitary time evolution operator.  In this section (\ref{f1}) is
expressed as the expectation value of a ``functional of paths'' with
respect to a complex probability on a space of paths
\cite{Muldowney}\cite{Katya_1}\cite{Katya_2}.

Following the construction in section \ref{sec:weyl}, irreducible pairs of
unitary operators, $U$ and $V$, can be constructed from $X$.  The
eigenvectors $\vert v_n \rangle$ of $V$ are identical to the
eigenvectors, $\vert n \rangle$,  of $X$, while the eigenvectors $\vert\bar{m} \rangle$ of
$U$ are complementary in the sense that
\beq
\vert \langle n \vert \bar{m} \rangle\vert^2 = {1 \over M}
\label{f2}
\eeq
for any $\vert n \rangle$ and $\vert \bar{m} \rangle$.
The operators $U$ and $V$ satisfy (\ref{s4}),(\ref{s5}),(\ref{s14}),(\ref{s16})
and
%$U$ satisfies
%\beq
%U \vert n \rangle = \vert n+1 \rangle
%\qquad
%U \vert \bar{n} \rangle = \vert \bar{n} \rangle e^{i {2\pi n \over M}},
%\label{f3}
%\eeq
%$V$ satisfies
%\beq
%V \vert \bar{n} \rangle = \vert \bar{n}-1 \rangle
%\qquad
%V \vert {n} \rangle = \vert {n} \rangle e^{i {2\pi n \over M}}
%\label{f4}
%\eeq
%and
\beq
VU = UV e^{i{2 \pi \over M} }.
\label{f5}
\eeq
The inner products of the eigenvectors of these complementary operators are
\beq
\langle n \vert \bar{m} \rangle ={e^{-i 2mn\pi/N} \over \sqrt{M}}.
\label{f6}
\eeq
%\[
%\sum_m \langle n \vert \bar{m} \rangle = \delta_{n0}\sqrt{M}.
%\]

The time-evolution operator can be expressed as
the $N^{th}$ power of a unitary transfer matrix, $T$:
\beq
e^{-iHt } = T^N \qquad T=e^{-iH \Delta t } \qquad \mbox{where} \qquad \Delta t= t/N .
\label{f7}
\eeq
Since  the pair of operators $U$ and $V$ is irreducible, 
$T$ or $H$ can be expressed as degree $(M-1)\times (M-1)$
polynomials in $U$ and $V$ using (\ref{s27}):
\beq
T= \sum_{mn} t_{mn} U^m V^n
\qquad H= \sum_{mn} h_{mn} U^m V^n .
\label{f8}
\eeq
The quantities $t_{mn}$ and $h_{mn}$ are complex valued functions of
two discrete variables.  The eigenvalues of $X$ can be taken as one of
the variables.  For the purpose of comparing to path integrals
expressed in terms of canonical coordinates and momenta it is useful
to define another Hermitian operator, $P$, that has the same
eigenvectors as $U$ with eigenvalues, $p_n$, analogous to the relation
between $X$ and $P$,
\beq
P \vert \bar{m} \rangle = p_m \vert \bar{m} \rangle .
\label{f9}
\eeq
The eigenvalues $x_n$ and $p_m$ will be referred to as
``coordinates'' and ``momenta'' for the purpose of illustration, although
in general they have no relation to coordinates and momenta.

In this section the choice of the eigenvalues, $p_m$, is not important.
All that is used in computations are the eigenvectors,
$\vert \bar{m} \rangle$.
Matrix elements of the transfer matrix in a mixed basis of eigenvectors
of $x$ and $p$
can be computed using (\ref{s4}),(\ref{s5}), (\ref{s16}) in (\ref{f8}):
\beq
\langle \bar{n} \vert T \vert {m} \rangle =
\langle \bar{n} \vert {m} \rangle \tilde{T}_{nm}
\label{f10}
\eeq
where the numerical coefficients, $\tilde{T}_{nm}$, are
\beq
\tilde{T}_{nm} =\sum_{kl}t_{kl} e^{- i\frac{2 \pi (nk-ml)}{M}}.
\label{f11}
\eeq
Mixed matrix elements of the Hamiltonian have a similar form 
\beq
\langle \bar{n} \vert H \vert {m} \rangle =
\langle \bar{n} \vert {m} \rangle \tilde{H}_{nm}
\qquad
\tilde{H}_{nm} =\sum_{kl}h_{kl} e^{- 2 \pi i {(nk-ml)\over M}}.
\label{f11a}
\eeq
Changing the ``momentum''
$\vert \bar{m} \rangle$ basis back to the  ``coordinate'' $\vert n \rangle$ basis gives
\beq
\langle k \vert T \vert m \rangle =
\sum_n\langle k \vert \bar{n} \rangle
\langle \bar{n} \vert {m} \rangle \tilde{T}_{nm} \qquad
\label{f12a}
\eeq
and
\beq
\langle k \vert H \vert m \rangle =
\sum_n\langle k \vert \bar{n} \rangle
\langle \bar{n} \vert {m} \rangle \tilde{H}_{nm} .
\label{f12b}
\eeq
Matrix elements of 
the time evolution
operator in the coordinate basis can be expressed,
using representation (\ref{f12a}-\ref{f12b}) in (\ref{f7}), as 
\[
\langle k_f \vert e^{-iHt} \vert k_i \rangle =\langle k_f \vert [e^{-iHt/N}]^N \vert k_i \rangle =
\]
\[
\sum_{k_1 \cdots k_N} \langle k_f \vert T \vert k_N \rangle
\langle k_N \vert T \vert k_{N-1} \rangle \cdots
\langle k_2 \vert T \vert k_1 \rangle
\langle k_1 \vert T \vert k_i \rangle =
\]
\[
\sum_{k_1,n_1 \cdots k_N, n_N}\langle k_f  \vert \bar{n}_N \rangle
\langle \bar{n}_N \vert {k}_N \rangle
\langle k_N  \vert \bar{n}_{N-1} \rangle
\langle \bar{n}_{N-1} \vert {k_{N-1}} \rangle \cdots 
\]
\[
\langle k_3  \vert \bar{n}_2 \rangle
\langle \bar{n}_2 \vert {k}_2 \rangle
\langle k_2 \vert \bar{n}_1 \rangle
\langle \bar{n}_1 \vert {k}_i \rangle \times
\]
\beq
\tilde{T}_{n_Nk_N}\tilde{T}_{n_{N-1}k_{N-1}}\cdots \tilde{T}_{n_2k_2}
\tilde{T}_{n_1k_i}.
\label{f13}
\eeq
The time step, $t/N$, appears in the coefficients $\tilde{T}_{n_mk_m}$.  These are complex numbers.

If the $ \tilde{T}_{n_mk_m}$ are all set to 1 then what remains after
performing the completeness sums is the overlap of the final
``coordinate'' with the initial ``coordinate'', $\langle k_f \vert k_i
\rangle =\delta_{k_f,k_i}$.  An additional sum over $k_i$ (resp $k_f$)
or gives 1, independent of $k_f$ (resp $k_i$). This motivates the
definition of the complex probability: 
\[
P_N(k_f; n_N,k_N,\cdots n_1,k_1):=
\]
\[
\langle k_f \vert \bar{n}_N \rangle \langle \bar{n}_N \vert {k}_N
\rangle \langle k_N \vert \bar{n}_{N_1} \rangle \langle \bar{n}_{N-1}
\vert {k_{N-1}} \rangle \times
\]
\[
\vdots
\]
\beq \langle k_3 \vert \bar{n}_2 \rangle \langle \bar{n}_2 \vert {k}_2
\rangle \langle k_2 \vert \bar{n}_1 \rangle \langle \bar{n}_1 \vert
        {k}_1 \rangle
\label{f14}
\eeq which by completeness satisfies \beq \sum_{n_1,k_1\cdots n_N, k_N}
P_N(k_f; n_N,k_N,\cdots n_1,k_1) =1
\label{f15}
\eeq
for any $\vert k_f \rangle $.
$P_N$ assigns a complex weight to a pair of eigenvalues of $X$ and $P$ at
each of $N$ time steps.
%for the possible
%values of the two non-commuting operators $U$ and
%$V$ (resp. $P$ and $X$) at each time step.

Computation of the transfer matrix is equivalent to a diagonailzation of the
Hamiltonian, but for sufficiently small time steps it can
be approximated without diagonalizing the Hamiltonian.
This approximation becomes exact in the limit that the size of the time steps vanish,
which is expressed using the Trotter product formula
\beq
e^{-iHt} = \lim_{N \to \infty}  (e^{-iHt/N})^N = 
\lim_{N \to \infty}  (I- iHt/N)^N
\label{f15b}
\eeq
where only the first-order term in time in the exponent contributes in the
$N \to \infty$ limit. 
This limit converges strongly to
$e^{-iHt}$  as $N\to \infty$ \cite{Simon}.
In the finite-dimensional case (\ref{f15b}) follows from
the identity $e^x= \lim_{N \to \infty} (1+\frac{x}{N})^N$. 

In the small $\Delta t= t/N$ limit 
\[
\langle \bar{m} \vert T \vert n \rangle \approx
\langle \bar{m} \vert I -i H \Delta t \vert n \rangle =
\]
\beq
\langle \bar{m} \vert n \rangle (1 -i \tilde{H}_{mn} \Delta t) \approx 
\langle \bar{m} \vert n \rangle e^{ -i \tilde{H}_{mn} \Delta t}.
\label{f16}
\eeq
In this expression $\tilde{H}_{mn}$ (\ref{f11a}) is an ordinary function of $p_m$ and $x_n$.

If the quantum Hamiltonian $H$ is expressed with the $P$ operators
to the right of the $X$ operators, $\tilde{H}_{mn}$ would be the Hamiltonian
with these operators replaced by the ``coordinates'' and ``momenta''.
It a discrete analog of a classical phase space Hamiltonian evaluated
at a point in phase space.

%It is instructive to make the following analogy with an ordinary path integral
%in the limit of small $\Delta t$.  In this case 
%$n_m\to p_m,k_m\to q_m$ are labels for
%classical coordinates and momenta at time
%$t$ (in this case they are discrete).
%\[
%\langle p_m \vert e^{-iH\Delta t} \vert q_m \rangle  \approx
%\]
%\[
%\langle p_m \vert (I -iH \Delta t \vert q_m \rangle \approx
%\]
%\[
%\langle p_m \vert q_m \rangle (1 -i h(p_m,q_m) \Delta t ) \approx
%\]
%\[
%\langle p_m \vert q_m \rangle (e^{-i h(p_m,q_m) \Delta t}
%\]
%where we assumed that commutation relations are used to order all of the
%momentum operators to the left of the coordinates.
It follows from (\ref{f10}) and (\ref{f16}) that
\[
\tilde{T}_{nk}\approx e^{-i \tilde{H}(p_{n},x_{k}) \Delta t}
\]
and 
\beq
\tilde{T}_{n_Nk_N}\tilde{T}_{n_{N-1}k_{N-1}}\cdots \tilde{T}_{n_2k_2}
\tilde{T}_{n_1k_i} \approx
e^{-i \sum_m \tilde{H}(p_{n_{m}},x_{k_{m}}) \Delta t},
\label{f17}
\eeq
which is a functional of a classical ``phase space'' path
$p(t),x(t)$ with endpoints at $x(0)$ and $x(t)$, where $p(t),x(t)$ can take
on $M$ discrete values, $p_n$ and $x_m$ at each time step.  These
``phase space paths'' are generally nowhere continuous.
This is classical in the sense
that even though the operators $P$ and $X$ do not commute, $\tilde{H}(p_n,x_m)$ is an ordinary
function of the eigenvalues $p_n$ and $x_m$.  
%
%\[
%e^{-i \int_0^t h(p(t'),q(t')) dt'}
%\]
%along a path with endpoint at $q(0)$ and $q(t)$.  With this interpretation
%the trotter product formula can be interpreted as the expectation of 
%\[
%e^{-t \int_0^t h(p(t'),q(t')dt'}
%\]
Following the definition in section \ref{sec:complexp}
a {\it cylinder set of paths} is the set of
all phase space paths,
$p(t)$ and $x(t)$ that take on one of the $M$ possible eigenvalues
of $P$ and $X$ at each of the $N$ time slices, subject to initial and final
values of the ``$x$'' variable.  With this definition
of cylinder set
\beq
P_N(k_f; n_N,k_N,\cdots n_1,k_1)
\label{f18}
\eeq
can be interpreted as a complex probability on the cylinder set
of ``phase space'' paths that start at $x_{k_1}$, end at $x_{k_f}$
and have values $p_{n_m},x_{k_m}$ at the $m^{th}$ time
slice.  This complex probability  satisfies (\ref{cp1}),
(\ref{cp2}) and (\ref{cp3}).

From a quantum mechanical point of view a $M$-valued path can be
thought of as an ordered sequence of quantum transition amplitudes
that alternate between eigenstates of irreducible pairs of observables.

Combining (\ref{f13}),(\ref{f14}) and (\ref{f16}) for a sufficiently large $N$
in the Trotter product formula 
gives the approximation
\[
%\boxed{
\langle k_f \vert e^{-iHt} \vert k_i \rangle \approx
\]
\beq
\sum_{n_N,k_N,\cdots n_1,k_1}
P_N(k_f; n_N,k_N,\cdots n_1,k_1)
e^{-i \sum_m \tilde{H}(x_{n_m} p_{k_m}) \Delta t}\delta_{k_1k_i}
%}
\label{f19}
\eeq
which converges in the limit that $N\to \infty$.
This approximates the transition probability amplitude as the
expectation value of a functional of the ``classical Hamiltonian''
$\tilde{H}(x,p)$ over cylinder sets of ``paths in phase space''.  Here
the ``coordinate'' paths have fixed endpoints at $0$ and $t$, while
the ``momentum'' paths are unconstrained.  In this case the complex
probability interpretation is a consequence of the completeness relation.

In general the number of cylinder sets that must be summed over is
prohibitively large, $(\sim M^{2N})$.  The definition (\ref{f14})
implies that the complex probability can be factored into a product of
one-step probabilities
\[
P_N(k_f; n_N,k_N,\cdots n_1,k_1)=
\prod_{i=1}^N \langle k_{i+1} \vert  \bar{n}_{i} \rangle
\langle \bar{n}_{i} \vert {k_{i}} \rangle :=
\]
\beq
\prod_i P(k_{i+1};n_i,k_i), \qquad k_{N+1}=k_f .
\label{f19a}
\eeq
Using (\ref{f19a}) the transition amplitude (\ref{f1}) can be approximated by
\[
\langle k_f \vert e^{-iHt} \vert k_i \rangle \approx
\]
\beq
\sum_{n_N,k_N,\cdots n_1,k_1}
\prod_i \left (P(k_{i+1}; n_i,k_i)
e^{-i \tilde{H}(x_{n_i} p_{k_i}) \Delta t}\right )\delta_{k_1k_i}
\label{f19b}
\eeq
where $k_{N+1}=k_f$.
This reduces the computation of (\ref{f19})  to computing the $N^{th}$ power of a $M\times M$ matrix.

In the discrete case time evolution can be solved exactly by diagonalizing
the $M\times M$ Hamiltonian matrix in the $x$ basis, however the appeal of the
discrete path integral is that it is a model of a quantum circuit.
Accuracy can be improved by using a higher-order approximation to the
transfer matrix at each step.   The complex probability interpretation is a
natural way to think about real-time path integrals for finite systems, since
concept of continuously parameterized paths that have $M$ possible
values does not really make sense.

The advantage of this approach is that the time evolution is
approximated by calculating the expectation value of a functional of a
finite set of cylinder sets of  ``classical paths'' with respect to a complex
probability distribution.  The ability to exactly factor the complex
probability into products of one-step probabilities for each time
step facilitates the computation.  Finally, exact unitarity is
maintained at each step. While the paths in each cylinder set are generally nowhere continuous,   the resulting amplitude is an entire function of time.
While the limit for continuous time evolution exists \cite{Katya_1}\cite{Katya_2}, real computations are truncated at a finite number of time steps.

%When the Hilbert space is two dimensional the difference in these two
%types of logic is encoded in bits or qubits respectively.  A bit can
%only be in one of two states, true or false.  A qubit is represented
%by a ray; if we ask ``does the state represented by this
%ray have spin up in the $z$ direction there are three possible
%outcomes: (1) the probability of measuring it to have spin up in
%the $z$ direction state is 1 (true), (2) the probability of measuring
%it to have spin up in the $z$ direction is 0 (false) and (3) the
%probability of measuring it to have spin up in the $z$ direction is
%strictly between $0$ and $1$.  Geometrically in case (1) the ray is in
%the spin up in the $z$ direction subspace, in case (2) the ray is
%orthogonal to the spin up in the $z$ direction subspace, and in case
%(3) the ray has a non-zero projection on both subspaces.

The discrete path integral developed in section (\ref{sec:findim}) represents
the transition probability amplitude as the expectation value of a
random variable on a space of sequences of transition amplitudes
labeled by paths in the space of eigenvalues of non-commuting 
observables.

%In quantum mechanics observable quantities are represented by linear
%operators $A$ on a Hilbert space.  The only possible outcomes for
%measuring $A$ are one of its eigenvalues, $a_n$ (this assumes $A$ is a
%normal operator whose eigenvectors form a basis). In this case the
%Hilbert space has a decomposition ${\cal H} = \oplus_n {\cal H}_n$,
%where the ${\cal H}_n$ are $A$-invariant subspaces of ${\cal H}$.
%
%The mean value of a measurement of $A$ in state $\vert b\rangle$ is  
%\beq
%\langle b \vert A \vert b \rangle =
%\sum_n P_{a_nb} a_n
%\qquad a_n \mbox{ eigenvalue of A} 
%\label{ql3}
%\eeq
%which is the weighted average of the quantum probabilities
%for a  measurement of $b$ to be in one of the eigenstates of $A$.

\section{Quantum logic}\label{sec:logic}

The interest in discrete path integrals is that they serve as models
of quantum computers.  The difference between classical and quantum
computers is related to the difference between classical and quantum
logic.  Classically statements have two possible outcomes; true or
false.  Different statements can be combined into compound statements
which have two possible outcomes.  The outcomes are determined by logical
truth tables,  which in turn can be represented by classical computer
gates.

In a quantum system the presence of non-commuting observables
complicates the logic.  Specifically the results of measurements can
depend on the order of the measurements.  The propositional calculus
behaves like properties of subspaces of a Hilbert space.  States in
quantum mechanics are identified with rays or one dimensional
subspaces.  Given another subspace in the Hilbert space there are
three possibilities: the ray is in the subspace, the ray is orthogonal
to the subspace, or the ray has a non-vanishing projection on the
subspace and its orthogonal complement.  Then the question ``will the
state always be measured to be in a particular invariant subspace of
some operator $A$'' is always, never, or with a probability $P$
satisfying $0<P<1$.  This feature can be represented using quantum
gates associated with non-commuting operators, which can be modeled
using irreducible sets of qubit gates.  As an illustration consider a system in a spin up state in the $z$ direction.  Measuring spin up in the $z$ direction will always give a positive result, measuring spin down in the $z$ direction will always give a negative result, but measuring spin up in the $x$ direction
will give a positive result 50 \% of the time.

One consequence of the presence of non-commuting operators,  
pointed out by Birkhoff and Von Neuman \cite{birkhoff}, was 
that the distributive laws of classical logic
\[
(A \mbox{ and } B) \mbox{ or } C = (A \mbox{ or } C) \mbox{ and }  (B \mbox{ or } C)
\]
\[
(A \mbox{ or } B) \mbox{ and } C = (A \mbox{ and } C ) \mbox{ or } (B \mbox{ and }  C)
\]
can be violated in quantum systems when the statements $A$, $B$ and $C$ are
associated with non-commuting operators. 

\section{Qubits}\label{sec:qubit}

One property of Schwinger's discrete Weyl algebra is that
it has a natural representation in terms of qubits.  When $M$ can be
factored into products of prime numbers the $U$ and $V$ operators
can be replaced
by an algebra of commuting pairs of operators with cycles the length
of each prime factor.  The case of most interest for quantum computing
is when $M=2^L$.  In that case the irreducible algebra is represented
by tensor products of qubit gates.  For systems where the number of degrees
of freedom $K$  is not a power of two, they can be embedded in a space of
dimension $M=2^L$, for $M < K$. 

Consider the case where  $M=2^L$ for large $L$.  The indices $n= 0 \cdots 2^L-1$
can be labeled by $L$ numbers that can only take the values $0$ and $1$.  It has
a L-bit representation
$n \leftrightarrow (n_1,n_2,\cdots, n_L)$
\beq
n=\sum_{m=1}^L n_m 2^{m-1}. 
\label{qb1}
\eeq
This correspondence results in the identifications
\beq
\vert u_{n_1\cdots n_L} \rangle :=\vert u_n \rangle
\qquad
\vert v_{n_1\cdots n_L} \rangle :=\vert v_n \rangle .
\label{qb2}
\eeq
Define unitary operators $U_i$ and $V_i$ by
\beq
U_i \vert v_{n_1\cdots n_L} \rangle =
\vert v_{n_1\cdots [n_i+1]_{\mbox{mod\,2}} \cdots n_L} \rangle
\label{qb3}
\eeq
\beq
V_i \vert u_{n_1\cdots n_L} \rangle =
\vert u_{n_1\cdots [n_i-1]_{\mbox{mod\,2}} \cdots n_L} \rangle .
\label{qb4}
\eeq
Applying what was done in the general case to $M=2^L$ gives
\beq
U_i^2-1= V_i^2 -1=0,
\label{qb5}
\eeq
\beq
[U_i,U_j]=[V_i,V_j]=0 \qquad [U_i,V_j]=0 \qquad i \not=j.
\eeq
Equation (\ref{s23}) for $M=2$ gives
\beq
V_i U_i = U_i V_i e^{i \pi}.
\label{qb6}
\eeq
It follows from (\ref{qb1}) that  
\beq
U^n = \prod_{m=1}^L U_m^{n_m}
\label{qb7}
\eeq
\beq
V^n = \prod_{m=1}^L V_m^{n_m}.
\label{qb8}
\eeq
These equations can be understood by noting
\[
\vert v_n \rangle = U^n\vert v_{0} \rangle =
U_1^{n1} \cdots U_L^{n_L} \underbrace{\vert 0,\cdots , 0
\rangle}_{\mbox{ $L$ times}}.
\]
so each of the $2^L$ independent powers of $U$ (or $V$) can be expressed
as a product of $U_i^0=I$ or $U_i^1$ for $L$ different $U_i$ (or $V_i$).
%The starting point can be changed by letting $n = m+k$ and $n_i=m_i+k_i$.
Since $U$ and $V$ can be constructed from the $U_i$ and $V_i $ the
set of $\{ U_i\}$ and $\{ V_i\}$ is also irreducible.

A matrix representation of $U_i$ and $V_i$ acting on the $i$-th qubit is
given by the Pauli matrices $\sigma_1$ and $\sigma_3$:
\beq V_i = \sigma_3 \qquad U_i = \sigma_1
\label{qb9}
\eeq which are simple quantum gates.  In this representation,
$v_0=u_0=1; v_1=u_1=-1$ and \beq \vert v_0 \rangle = \left(
\begin{array}{c}
1 \\ 0\\
\end{array}
\right ) \qquad \vert v_1 \rangle = \left (
\begin{array}{c}
0 \\ 1\\
\end{array}
\right )
\label{qb10}
\eeq \beq \vert u_0 \rangle ={1 \over \sqrt{2}} \left(
\begin{array}{c}
1 \\ 1\\
\end{array}
\right ) \qquad \vert u_1 \rangle = {1 \over \sqrt{2}} \left (
\begin{array}{c}
1 \\ -1\\
\end{array}
\right ).
\label{qb10}
\eeq The operators $\sigma_1$ and $\sigma_3$ \beq U_i= \sigma_1 =
\left (
\begin{array}{cc}
  0&1\\ 1&0\\
\end{array}
\right ) \qquad V_i= \sigma_3 = \left (
\begin{array}{cc}
  1&0\\ 0&-1\\
\end{array}
\right ) \qquad
\label{qb11}
\eeq satisfy (\ref{s2}) and (\ref{s14}) for $M=2$:
\[
U_i \vert v_0 \rangle \vert v_1 \rangle \qquad U_i \vert v_1 \rangle
\vert v_0 \rangle
\]
\[
V_i \vert u_0 \rangle \vert v_1 \rangle \qquad V_i \vert u_1 \rangle
\vert v_0 \rangle .
\]
They also satisfy \beq \sigma_3 \sigma_1 = \sigma_1 \sigma_3 e^{{2\pi
    i \over 2}} \qquad (\sigma_1^2 -1)= (\sigma_3^2-1) =0 .
\label{qb12}
\eeq Any linear operator $A$ on this 2-dimensional vector space is a
polynomial with constant coefficients $a_i$ in these operators: \beq A
= a_1I + a_2 \sigma_1 + a_3 \sigma_3 +a_4 \sigma_3\sigma_1 .
\label{qb13}
\eeq In this case the Hilbert space is represented by $L$ qubits. The
irreducible set of operators, $U_i$ and $V_i$ are represented by $L$
pairs of Pauli matrices ($\sigma_1$ and $\sigma_3$) that act on each
qubit. This representation has the advantage that it is local in the
sense that the $U_i$ and $V_i$ operators act on a single qubit
and the operators that act on different qubits commute.
Equations (\ref{qb7}-\ref{qb8}) relate the operators that appear in
the discrete path integral to tensor products of single qubit operators
$\sigma_1$ and $\sigma_3$.  The irreducibility of the $U_i,V_i$ operators
implies that any gates involving several qubits can be expressed
as products of these elementary gates.  The advantage of these
single qubit gates is that they are efficient at representing more
complicated gates that involve a limited number of degrees of freedom.

%Note for $M$ odd the same construction works with
%$3\times 3$ matrices with
%\beq
%U_i = \left (
%\begin{array}{ccc}
%  0&0&1\\
%  1&0&0\\
%  0&1&0
%\end{array}
%\right )
%\qquad
%V_i = \left (
%\begin{array}{ccc}
%1&0&0\\
%0&e^{2 \pi i/3}&  0\\
%0 & 0 & e^{4 \pi i /3}
%\end{array}
%\right ).
%\label{qb14}
%\eeq

\section{Schwinger's continuum limit}\label{cont}

While quantum computers are discrete quantum systems, many problems of
interest involve observables like momenta, coordinates, and canonical
fields that have continuous spectra. Applications require discrete
approximations to these continuous systems.  It is possible to use the
discrete algebra generated by $U$ and $V$ to make a discrete
approximation to the continuum in the large $M$ limit. The limit is
designed to give a representation of the Weyl algebra.

For large $M$ Schwinger \cite{schwinger} defines the small quantity
$\epsilon$ by
\beq
%\boxed{
\epsilon^2 :=
{2 \pi /M}.
%}
\label{cl1}
\eeq
For the purpose of approximating the continuum it is convenient
(but not necessary) to
choose $M=2K+1$ odd and number the eigenvectors and eigenvalues from
$-K \leq n \leq K$ instead of $0$ to $M-1$ or $1$ to $M$ (For even $M$
the indices could be labeled $-M/2+1 \leq n \leq M/2$).
Discrete approximations to continuous
variables $p$ and $x$ are defined by
\beq
p_l = l \epsilon = l \sqrt{2 \pi \over M}   \qquad x_l = l \epsilon = 
l \sqrt{2 \pi \over M} \qquad  -K\epsilon \leq x_l,p_l \leq K\epsilon
\label{cl2}
\eeq
where
\beq
K\epsilon  = \sqrt{M\pi \over 2} - \sqrt{\pi \over 2M}.
\label{cl3}
\eeq
With these definitions the separation between successive values of $p_l$
and $x_l$, $p_{l+1}-p_l =x_{l+1}-x_l = \epsilon$
vanishes as $M \to \infty$ while at the
same time the maximum and minimum values of $p_l$ and $x_l$,
$p_{\pm K} = x_{\pm K} = \pm (
\sqrt{M\pi \over 2} - \sqrt{\pi \over 2M} )$ approach $\pm\infty$ in
same limit.

While for finite $M$ any vector with a finite number of elements has a
finite norm, in the continuum limit ($M\to \infty$) this is no longer
true so the limiting vectors with finite norm should be square
summable.  This means that components of vectors with large $\vert l
\vert$ should approach $0$ in the $M\to \infty$ limit.  

For unitary operators $U$ and $V$ given by (\ref{s2}) and (\ref{s14})
Hermitian operators $\tilde{p}$ and $\tilde{x}$ are defined by
\beq
%\boxed{
V = e^{i \epsilon \tilde{p}}
\qquad
U= e^{i \epsilon \tilde{x}}.
%} .
\label{cl4}
\eeq
These can be used to define
\beq
V(x_m) = e^{i\tilde{p} x_m} = e^{i\tilde{p} \epsilon m} = V^m
\label{cl5}
\eeq
\beq
U(p_n) = e^{i\tilde{x} p_n} = e^{i\tilde{x} \epsilon n} = U^n .
\label{cl6}
\eeq
With definitions (\ref{cl5}) and (\ref{cl6})
equation (\ref{s23}) becomes
\[
V(x_m)U(p_k)= U(p_k)V(x_m) e^{i2 \pi mk \over M}=
\]
\beq
U(p_k)V(x_m) e^{i \epsilon m \epsilon k}=
U(p_k)V(x_m)e^{ip_kx_m}
\label{cl7}
\eeq
%\beq
%\boxed{
%V(x_m)U(p_k)= 
%U(p_k)V(x_m)e^{ip_kx_m}
%}
%\label{cl8}
%\eeq
which is the Weyl \cite{Weyl} (exponential) form of the canonical commutation
relations, where in this
case the variables are discrete.
In order to take the continuum limit define numbers $p:=\epsilon n$ and
$x= \epsilon m$.  This motivates the definitions 
%Equations (\ref{cl5}-\ref{cl6}) motivate the definitions
\beq
%\boxed{
dp = \epsilon dn = \sqrt{2 \pi \over M} dn
\qquad
dx = \epsilon dm = \sqrt{2 \pi \over M} dm 
%}
\label{cl9}
\eeq
and
\beq
\int dp \approx \sum_{n=-K}^K {dp \over dn} = \epsilon \sum_{n=-K}^K
\eeq
\beq
\int dx \approx \sum_{n=-K}^K {dx \over dm} = \epsilon \sum_{m=-K}^K
\label{cl9}
\eeq
which becomes a Riemann integral in the limit that $M\to \infty$.
It follows from (\ref{cl4}) that eigenvectors of $V$ are also
eigenvectors of $\tilde{p}$ and the eigenvectors of $U$ are also
eigenvectors of $\tilde{x}$.  Choosing the normalization of the states $\vert
p_n\rangle$ and $\vert x_n \rangle$ so
\beq
I = \sum_{l=-K}^K  \vert v_l \rangle \langle v_l \vert =
\sum_{l=-K}^K  \vert p_l \rangle dp_l \langle p_l \vert =
\sum_{l=-K}^K  \vert p_l \rangle \epsilon \langle p_l \vert
\label{cl10}
\eeq
\beq
I = \sum_{l=-K}^K  \vert u_l \rangle \langle u_l \vert =
\sum_{l=-K}^K  \vert x_l \rangle dx_l \langle x_l \vert =
\sum_{l=-K}^K  \vert x_l \rangle \epsilon \langle x_l \vert , 
\label{cl11}
\eeq
which for finite $\epsilon$ approximates the sum over small steps in $p$ or $x$ as integrals.
Equations (\ref{cl10}) and (\ref{cl11})
imply the relations
\beq
%\boxed{
\vert p_l \rangle := \vert v_l \rangle /\sqrt{\epsilon}
%}
\label{cl12}
\eeq
and 
\beq
%\boxed{
\vert x_l \rangle := \vert u_l \rangle /\sqrt{\epsilon} .
%}.
\label{cl13}
\eeq
Using these identities gives
\beq
\langle p_m\vert x_n \rangle = 
        {1 \over \epsilon} \langle v_m \vert u_n \rangle =
        {1 \over \epsilon \sqrt{M}}e^{-i\frac{2 \pi  mn}{ M}}
        =
        {1 \over \sqrt{2 \pi}} e^{-i p_m x_n}
\eeq
\beq
\langle p_m\vert p_n \rangle = 
{1 \over \epsilon} \langle v_m \vert v_n \rangle =
{\delta_{mn} \over \epsilon}
\label{cl14}
\eeq
and
\beq
\langle x_m\vert x_n \rangle = 
{1 \over \epsilon} \langle u_m \vert u_n \rangle =
{\delta_{mn} \over \epsilon}.
\label{cl15}
\eeq
In this notation equations (\ref{cl5}) and (\ref{cl6})
with (\ref{s2}) and (\ref{s14}) give 
\beq
U(x_m) \vert x_n \rangle = \vert x_m+x_n \rangle
\label{cl16}
\eeq
\beq
V(p_m) \vert p_n \rangle = \vert p_m-p_n \rangle
\label{cl16}
\eeq
which can be expressed in terms of the ``continuum variables'' as  
\beq
U(x) \vert x' \rangle = \vert x'+x \rangle
\label{cl16}
\eeq
\beq
V(p) \vert p' \rangle = \vert p'-p \rangle.
\label{cl16}
\eeq

\section{Complex probabilities in real-time path integrals on infinite
dimensional Hilbert spaces}\label{path}

%The path integral for a system with one continuous degree of freedom is
%formulated using the discrete representation discussed section \ref{weyl}.
%Following references \cite{Muldowney}\cite{Katya_1}\cite{Katya_2},
%the path integral will be defined as the expectation value of a
%potential functional with respect to a complex probability
%distribution.
%
%To do this it is necessary to:
%\begin{itemize}
%\item[1.)] Define the space of paths
%\item[2.)] Define complex probabilities on the space of paths
%\item[3.)] Identify the path integral with the expectation value of a
%functional on the space of paths.
%\end{itemize}
%
For problems involving scattering or canonical field theories the
relevant operators have continuous spectra.  Systems with continuous
variables can be approximated by discrete systems following section
\ref{cont}.  In the continuous case a general Hamiltonian, $H(p,x)$, can be
approximated by
\[
H = \int \tilde{H}(p,x) U(x) V(p) dxdp \approx
\]
\beq
\sum_{ij}\tilde{H}(p_j,x_i) U(x_i)
V(p_j) dx_i dp_j
\label{cpi1}
\eeq
with a similar representation for the transfer matrix
\[
T := e^{-iH\Delta t}=\int \tilde{T}(x,p) U(x) V(p) dxdp \approx
\]
\beq
\sum_{ij}\tilde{T}(x_i,p_j) U(x_i)V(p_j) dx_i dp_j
\label{cpi2}
\eeq
where
\beq
\langle p \vert T \vert x \rangle = \tilde{T}(p,x)\langle p \vert x \rangle
\qquad
\mbox{and}
\qquad
\langle p \vert H \vert x \rangle = \tilde{H}(p,x)\langle p \vert x \rangle .
\label{cpi3}
\eeq
Here the quantities with the ``tildes'' are matrix elements of operators
in a mixed basis 
where the canonical
commutation relations are used to order the momentum operator to the left of the coordinate operators.

The probability amplitude $\langle x_f \vert e^{-iHt} \vert x_i \rangle$ can be expressed in terms of transfer matrices, $T:=e^{-iHt/N}$,
\beq
\langle x_f \vert e^{-iHt} \vert x_i \rangle =
\langle x_f \vert (e^{-iHt/N})^N \vert x_i \rangle =
\langle x_f \vert T^N \vert x_i \rangle .
\label{cpi5}
\eeq
Inserting intermediate states gives
\[
\langle x_f \vert e^{-iHt} \vert x_i \rangle =
\]
\[
\int \prod_{i} dp_{i} dx_{i} \langle x_f \vert p_N \rangle \langle p_N \vert
T \vert x_N \rangle  \langle x_N \vert p_{N-1} \rangle \langle p_{N-1} \vert T \vert x_{N-1} \rangle \cdots
\]
\[
\langle x_3 \vert p_2 \rangle \langle p_2 \vert T \vert x_2 \rangle  \langle x_2 \vert p_{1} \rangle \langle p_{1} \vert T \vert x_{1} \rangle \langle x_1 \vert x_i \rangle = 
 \]
\[
\int \prod_{i} dp_{i} dx_{i} \langle x_f \vert p_N \rangle \langle p_N \vert x_N \rangle \tilde{T}(p_N,x_N) \langle x_N \vert p_{N-1} \rangle
\times
\]
\[
\langle p_{N-1} \vert q_{N-1} \rangle  \tilde{T}(p_{N-1},x_{N-1})\cdots
\]
\beq
\langle x_3 \vert p_2 \rangle \langle p_2  \vert x_2 \rangle \tilde{T}(p_2,x_2) \langle x_2 \vert p_{1} \rangle \langle p_{1} \vert x_{1} \rangle \tilde{T}(p_1,x_1) \langle x_1 \vert x_i \rangle .
\label{cpi6}
\eeq
This is exactly $\langle x_f \vert (e^{-i\frac{Ht}{N}})^N
\vert x_i \rangle = \langle x_f \vert e^{-i{Ht}}
\vert x_i \rangle$.  The Trotter limit justifies the replacement
\beq
\tilde{T}(p_i,x_i) \to e^{-i \tilde{H}(p_i,x_i) t/N}
\label{cpi7}
\eeq
which becomes exact in the limit $N \to \infty$.

Completeness implies that
\[
P(x_f;p_N,x_N, \cdots , p_1,x_1):=
\]
\[
\langle x_f \vert p_N \rangle
\langle p_N \vert x_N \rangle \langle x_N \vert p_{N-1} \rangle
\langle p_{N-1} \vert x_{N-1} \rangle \cdots
\]
\beq
\langle x_3 \vert p_2 \rangle \langle p_2 \vert x_2 \rangle  \langle x_2 \vert p_{1} \rangle \langle p_{1} \vert x_{1} \rangle.
\label{cpi8}
\eeq
satisfies
\beq
\int \prod_{i=1}^N dp_i dx_i P(x_f;p_N,x_N, \cdots , p_1,x_1) =1
\label{cpi9}
\eeq
independent of $N$. With the replacement (\ref{cpi7}) equation (\ref{cpi6}) becomes
\[
\langle x_f \vert e^{-iHt} \vert x_i \rangle =
\]
\[
\lim_{N\to \infty} \int \prod_{i} dp_{i} dx_{i}
P(x_f;p_N,x_N, \cdots , p_1,x_1)\times
\]
\beq
e^{-i \sum \tilde{H}(p_j,x_j) t/N}\delta (x_1-x_i)
\label{cpi10}
\eeq
where the limit is interpreted as a strong limit;
the initial coordinate must be multiplied by a wave packet and integrated.
The complex probability
interpretation follows by considering the integrals as Henstock integrals,
which are limits of generalized Riemann sums. Cylinder sets are
defined by sets of paths that go through a generalized Riemann interval at each time step.
The complex probability
admits a
factorization 
\[
\prod_{i=1}^N dp_i dx_i
P(x_f;p_N,x_N, \cdots , p_1,x_1):=
\]
\[
P(x_f;x_N,p_N)dx_N dp_N 
P(x_N;x_{N-1},p_{N-1})dx_{N-1} dp_{N-1} 
\cdots
\]
\beq
P(x_2;x_{1},p_{1})dx_{1} dp_{1} 
\label{cpi11}
\eeq
which expresses the complex probability as a product of one step
probabilities for each time step.  Using this factorization gives
\[
\langle x_f \vert e^{-iHt} \vert x_i \rangle =
\]
\[
\lim_{N\to \infty} \int \prod_{i} dp_{i} dx_{i} P(x_{i+1};p_i,x_i,)e^{-i \tilde{H}(p_i,x_i) t/N}\delta (x_1-x_i),
\]
\beq 
x_{N+1}=x_f
\label{cpi12}
\eeq

In the discrete approximation of section \ref{cont} the integrals are
replaced by sums over the discrete values of $p_m= m\epsilon$,
$x_m=m\epsilon$, $dp_m=\epsilon=dp$ and $dx_m=\epsilon=dx$ where $-K
\leq m \leq K$, $M=2K+1$ and $\epsilon^2 = {2 \pi \over M}$ .
Equation (\ref{cpi9}) is still satisfied independent of the number $M$
of discrete values of $p_m$, $x_m$.  In the discrete case a cylinder
set is the set of paths that take on specific discrete values of $p=n\epsilon$
and $q=m\epsilon$ at each of the $N$ intermediate time steps. In that case the
complex probability for phase space paths becomes
\[
P(x_f;p_N,x_N, \cdots ,
p_1,x_1)\prod_i dp_idx_i  \to
\]
\[
P(x_f;p_{Ni_N},x_{Nj_N}, \cdots , p_{1i_1},x_{1j_1})\epsilon^{2N}:=
\]
\[
\langle x_f \vert p_{Ni_N} \rangle \epsilon
\langle p_{Ni_N} \vert x_{Nj_N} \rangle \epsilon\langle x_{Nj_N} \vert p_{N-1i_{N-1}}
\rangle \epsilon
\langle p_{N-1i_{N-1}} \vert x_{N-1 j_{N-1}} \rangle \epsilon \cdots
\]
\beq
\epsilon
\langle x_{3j_3} \vert p_{2i_2} \rangle
\epsilon
\langle p_{2i_2} \vert x_{2j_2} \rangle \epsilon  \langle x_{2j_2} \vert p_{1i_1} \rangle
\epsilon\langle p_{1i_1} \vert x_{1j_1} \rangle \epsilon.
\label{cpi13}
\eeq
Here these indices represent the discrete momenta and coordinates that define a path
in phase space.
This has the property that the sum over all of the $M^N$ cylinder sets for
$N$ intermediate time steps gives 1 independent of $x_f$. In this case there
are cylinder sets of paths in both the $p$ and $x$ variables. These are considered as
discrete approximations
to the phase space paths, $p(t),x(t)$.

In the Trotter limit the discrete transfer matrix and discrete
Hamiltonian (see (\ref{cpi1})),
\beq
H= \sum \tilde{H}(p_m,x_n)U(x_n)V(p_m)
\label{cpi14a}
\eeq
\beq
T = \sum \tilde{T}(p_m,x_n)U(x_n)V(p_m)
\label{cpi14b}
\eeq
are related by
\beq
\tilde{T}(p_m,x_n) \approx e^{-i  \tilde{H}(p_m,x_n) \Delta t},
\label{cpi15}
\eeq
where this approximation preserves unitarity.

In this case 
\[
\langle
x_f \vert e^{-i H t} \vert  x_i \rangle =
\]
\[
\lim_{N\to \infty} \sum
P(x_f;p_{Ni_N},x_{Nj_N}, \cdots , p_{1i_1},x_{1j_1}) \times
\]
\beq
e^{-i \sum \tilde{H}( p_{1i_k},x_{1j_k})t/N} \delta_{x_1,x_i}
\label{cpi16}
\eeq
which expresses time evolution as the expectation of a complex probability on
cylinder sets of paths in phase space. As in the continuous case the
complex probability factors into a product of $N$ one-step complex
probabilities:
\[
P(x_f;p_{=Ni_N},x_{Nj_N}, \cdots , p_{1i_1},x_{1j_1})=
\]
\beq
\prod_k [\langle x_{k+1i_{k+1}} \vert p_{ki_k} \rangle \epsilon 
\langle p_{ki_k} \vert x_{ki_k} \rangle \epsilon] \qquad x_{N+1}=x_f
\label{cpi17}
\eeq
and the probability amplitude for a transition from $x_i$ to $x_f$ after time $t$ becomes
\[
\langle
x_f \vert e^{-i H t} \vert  x_i \rangle =
\]
\beq
\lim_{N\to \infty}
\sum
\prod_k [\langle x_{k+1i_{k+1}} \vert p_{ki_k} \rangle \epsilon 
\langle p_{ki_k} \vert x_{ki_k} \rangle \epsilon e^{-i \tilde{H}(p_{ki_k} x_{ki_k})t/N}]
\delta_{x_1,x_i}
 \label{cpi18}
\eeq
where the sum is over cylinder sets.
This reduces the calculation to computing powers of a $M \times M$ matrix

For the examples discussed in sections (\ref{scatt}) and (\ref{qft})
$H$ has the form 
\beq
H(p,x) = H_1(p) + H_2(x) \qquad H_1(0) =0 
\label{cpi19}
\eeq
with
%\beq
\[
\tilde{H}(p,x) = \tilde{H}_1(p) + \tilde{H}_2(x) \qquad \tilde{H}_1(0) =0 .
\]
%\label{cpi19}
%\eeq
This is discussed in detail in (\ref{app1}).
In this case equation  (\ref{cpi18})  becomes
\[
\langle
x_f \vert e^{-i H t} \vert  x_i \rangle =
\]
\[
\lim_{N\to \infty}
\sum
\prod_k [\langle x_{k+1i_{k+1}} \vert p_{ki_k} \rangle \epsilon e^{-i \tilde{H}_1(p_{ki_k}) t/N}\langle p_{ki_k} \vert x_{ki_k} \rangle \times
\]
\beq
e^{-i \tilde{H}_2(x_{ki_k}) t/N}\epsilon ]\delta_{x_1,x_i}.
\label{cpi20}
\eeq
Since everything is finite the sums over the $p$ values can be computed
first, defining 
\[
P_X(x_f;x_{Ni_{N}},x_{N-1i_{N-1}}\cdots x_{1i_1}):=
\]
\beq
\sum_{p}
\prod_k \langle x_{k+1i_{k+1}} \vert p_{ki_k} \rangle e^{-i \tilde{H}_1(p_{ki_k}) t/N}
\epsilon \langle p_{ki_k} \vert x_{ki_k} \rangle \epsilon 
\label{cp21}
\eeq
which is complex valued function on cylinder sets of paths in the ``$x$'' variable.  It follows from
(\ref{f21}), that summing over
all of the cylinder sets, starting with the right most index and working to the left gives
\beq
\sum_{x}    P_X(x_f;x_{Ni_{N}},x_{N-1i_{N-1}}\cdots x_{1i_1})=
e^{-iH_1(0)t} =1.
\label{cpi22}
\eeq

With this definition
\[
\langle
x_f \vert e^{-i H t} \vert  x_i \rangle =
\]
\beq
\lim_{N\to \infty} \sum_q
P_X(x_f,x_{Nj_N}, \cdots ,x_{1j_1})
e^{-i \sum \tilde{H}_2(x_{1j_k})t/N} \delta_{x_1,x_i}.
\label{cpi23}
\eeq
As in the general case the complex probability factors into products of
conditional probabilities associated with each time step:
\beq
P_X(x_f;x_{Nj_N}, \cdots ,x_{1j_1})= \prod_k
P_X(x_{k+1i_{k+1}};x_{ki_k})
\label{cpi24}
\eeq
where
\[
P_X(x_{k+1i_{k+1}};x_{ki_k})=
\]
\beq
\sum_{p_ki_k} \langle x_{k+1i_{k+1}} \vert p_{ki_k} \rangle e^{-i \tilde{H}_1(p_{ki_k}) t/N}\epsilon
\langle p_{ki_k} \vert x_{ki_k} \rangle \epsilon
\label{cpi25}
\eeq
which gives the following expression for the probability amplitude
\[
\langle
x_f \vert e^{-i H t} \vert  x_i \rangle =
\]
\beq
\lim_{N\to \infty}
\sum
\prod_k [P_X(x_{k+1i_{k+1}};x_{ki_k})e^{-i \tilde{H}_2(x_{ki_k}t/N}]
\delta_{x_1 x_i}.
\label{cpi26}
\eeq
This expresses the transition amplitude as the $N^{th}$ power of a
$M \times M$ matrix.

Note that treating an infinite dimensional system as the limit of
finite dimensional systems rather than discretizing the infinite
dimensional system has some advantages.  Discretizing was used in
\cite{polyzou} where the rate of convergence was sensitive to how the
points used to evaluate the interaction in each interval were chosen.
The factorization of the complex probability into products of
conditional probabilities becomes approximate upon
discretization. These choices could in principle affect unitarity
numerically.  These issues do not appear in the discrete case.  In
addition, the discrete analogs of the Fresnel integrals are
well-defined finite sums.  The complex probability interpretation arises
naturally from the completeness relation.

\section{Scattering in the discrete representation}\label{scatt}

Formal scattering theory is an idealization.  A real scattering experiment
takes place in a finite volume during a finite time interval.  The relevant
physics is dominated by a finite number of degrees of freedom that are
limited by the energy and scattering volume.

The fundamental quantum mechanical observable is the probability
for a transition from a prepared initial state to a detected final state
\beq
P_{fi} = \vert \langle \psi_f (t)\vert \psi_i (t) \rangle \vert^2.
\label{sca1}
\eeq
While the individual states depend on time, the probability
(\ref{sca1}) is independent of $t$ due to the unitarity of the time
evolution operator.  The important constraint is that both states have to be
evaluated at the same time.  The problem of scattering theory is that
there is no common time when both the initial and final states are
simple.  On the other hand the initial state is simple before the
collision and the final state is simple after the collision.

The initial and final states at the time of collision can be
determined by evolving them from times where they behave like
non-interacting subsystems to the collision time.  Since localized
wave packets spread, the effects of spreading can be eliminated by
starting with localized wave packets at the collision time, evolving
them beyond the range of interactions using free time evolution, and
then evolving them back to the interaction region using the full
Hamiltonian.  The result is a unitary mapping that transforms the free
wave packet at the collision time to the dynamical wave packet at the
same time.

For scattering the Hamiltonian $H$ is the sum of a free Hamiltonian
$H_0$ and an interaction $V$.  In this section $U(t)$, $V$,
$T(z)$ represent the unitary time translation operator, the potential,
and the transition operator rather than the Weyl operators and the
transfer matrix.

If $U_0(t)$ and $U(t)$ represent the free and dynamical unitary time
evolution operators, $e^{-iHt}$ and $e^{-iH_0t}$,
then assuming the time of collision is
approximately at time $t=0$ the scattering asymptotic conditions have
the form
\beq
\Vert  U(\pm \tau)\vert \psi_{\pm}(0) \rangle -
U_0(\pm \tau) \vert  \psi_{0\pm}(0) \rangle \Vert 
\approx 0
\label{sca2}
\eeq
where the time $\tau$ is sufficiently large for the interacting particles to be
separated beyond the range of their mutual interactions.  This
expression is independent of $\tau$ for sufficiently large $\tau$, but
the minimum value of $\tau$ depends on the range of the interaction,
$V$, 
and the structure of $\vert \psi_{0\pm}(0) \rangle$.  Normally the dependence
on these conditions is removed by taking the limit $\tau \to
\infty$.  In this work, for computational reasons, it is desirable to
choose $\tau$ as small as possible, which requires paying attention to
the range of the interaction and the structure of the initial and
final states.

The unitarity of the time evolution operator means that (\ref{sca2})
can be replaced by
\beq
\Vert  \psi_{\pm} (0)\rangle -
U(\mp \tau)U_0(\pm \tau) \vert \psi_{0\pm}(0) \rangle \Vert 
\approx 0 
\label{sca3}
\eeq
for sufficiently large $\tau$.
The operators
\beq
\Omega_{\pm}(\tau) := U(\pm \tau)U_0(\mp \tau)
\label{sca4}
\eeq  
are unitary mappings from $\vert \psi_{0\pm} (0)\rangle$ to
$\vert \psi_{\pm}(0) \rangle$.

Using these definitions the scattering probability can be expressed as
\beq
P_{fi} = \vert \langle \psi_{+}(0)\vert \psi_{-}(0) \rangle \vert^2
=
\vert \langle \psi_{0+}(0)\vert  S(\tau) \vert \psi_{0-}(0) \rangle \vert^2 \label{sca5}
\eeq
where
\beq
S(\tau) := \Omega^{\dagger}(\tau)\Omega (-\tau)
\label{sca6}
\eeq
is the scattering operator.
Since $S(\tau)$ is unitary it can be expressed in terms of a self-adjoint
phase shift operator
\beq
S(\tau) = e^{2i \delta (\tau)}
\label{sca6}
\eeq
where $S(\tau)$ should be independent of $\tau$ for sufficiently
large $\tau$.

In a real experimental measurement the probability (\ref{sca5})
depends on the structure of the initial and final wave packets, which
cannot be precisely controlled by experiment.  If the matrix elements
of $S(\tau)$ in sharp-momentum states are slowly varying functions of
momentum, then the dependence on the wave packet factors out
\cite{brenig} and can be eliminated to compute differential cross
sections.  In this case the sharp-momentum matrix elements can be
approximated from the matrix elements using Gaussian (minimal
uncertainty) wave packets with a ``delta-function normalization'' that
are sharply peaked about the desired momenta.

%In the large $\tau$ limit sharp momentum matrix elements of $S(\tau)$
%exactly conserves energy; in a calculation using finite $\tau$ energy
%conservation is only approximate - so states should be chosen so they
%have the same mean energy.

This formulation of scattering admits a path integral
treatment.  As previously discussed scattering reactions are dominated
by a finite number of degrees of freedom.  The use of the discrete
Weyl representation has the advantage that unitarity is exactly
preserved on truncation to a finite number of degrees of freedom.
Alternative path integral treatments of scattering appear in
\cite{Campbell}\cite{Rosenfelder1}\cite{Rosenfelder2}.

The advantage of the discrete representation is that 
$U_0(-\tau)U(2\tau)U^0(-\tau)$ can be expressed as the limit of
products of the transfer matrices defined in section \ref{cont}
(see (\ref{cpi23}),(\ref{cpi26}))
\beq
S(\tau) = \lim_{N\to \infty} Y^{-N}X^{2N} Y^{-N}
\label{sca7}
\eeq
where
\beq
Y_{ij}= P_X(x_i,x_j,\Delta t)
\qquad
(X)_{ij}= P_X(x_i,x_j,\Delta t) e^{-i V(x_j)\Delta t},
\label{sca8}
\eeq
$\Delta t = \tau/N$ and $N$ is the number of Trotter time slices. 
Note also that
\beq
Y^{-N} = P_X(x_f,x_i,-N\Delta t).
\label{sca8a}
\eeq

Sharp-momentum matrix elements of the scattering operator can be expressed in
terms of the matrix elements of the transition operator,
\[
T(z) = V + V(z-H)^{-1}V
\]
\beq
\langle p_f \vert S \vert p_i \rangle =
\langle p_f \vert p_i \rangle
- 2 \pi i \delta (E_f -E_i) \langle p_f \vert T(E+i \epsilon)
\vert p_i \rangle
\label{sca9}
\eeq
where $T$ is approximately given by \cite{Brenig:1959}
\beq
T(E+i \epsilon) \approx V \Omega (-\tau)
\label{sca10}
\eeq
when evaluated in normalizable states with sharply peaked momenta.
This becomes exact in the limit $\tau \to \infty$.  The advantage of
this representation is that for scattering problems the
interaction $V$ is a short range operator that provides a volume
cutoff.

In the discrete representation sharp-momentum eigenstates are
normalizable however {\it they cannot be used in scattering calculations}
because they are completely delocalized in space, since the
discrete momenta and coordinates are complementary operators
- making it impossible to get to the asymptotic region.

The most straightforward way to construct suitable initial or final wave packets
in the discrete representation is to approximate the corresponding
minimal uncertainty states of the continuum theory.  The quantities
to control are the mean position,  momentum and the uncertainty
in both of these quantities defined for a given state $\vert \psi \rangle$
by:
\beq
\langle x \rangle_{\psi} := \sum_{n=-K}^K {\langle \psi \vert u_n \rangle
  n \epsilon \langle u_n \vert \psi \rangle \over
  \langle \psi \vert \psi \rangle }
\eeq
\beq
\langle p \rangle_{\psi} := \sum_{n=-K}^K {\langle \psi \vert v_n \rangle
n \epsilon \langle v_n \vert \psi \rangle \over
  \langle \psi \vert \psi \rangle }
%\label{sca12}
\eeq
\[
(\Delta x)^2 =\langle \psi \vert (x-\langle x \rangle)^2 \vert \psi \rangle =
\]
\beq
\sum_{n=-K}^K {\langle \psi \vert u_n \rangle
((n \epsilon)^2 - \langle x \rangle^2 )\langle u_n \vert \psi \rangle \over
  \langle \psi \vert \psi \rangle }
\label{sca13}
\eeq
 \[
(\Delta p)^2 = \langle \psi \vert (p-\langle p \rangle)^2 \vert \psi \rangle=
\]
\beq
\sum_{n=-K}^K {\langle \psi \vert v_n \rangle
((n \epsilon)^2 - \langle p \rangle^2 )\langle v_n \vert \psi \rangle \over
\langle \psi \vert \psi \rangle }.
\label{sca14}
\eeq
The continuum delta-function normalized minimal uncertainty states are
\beq
\langle p \vert \psi_{0}  (0) \rangle =
{1 \over 2 \sqrt{\pi} \Delta p} e^{- {(p-\langle p \rangle )^2 \over 4 (\Delta p)^2}} .
\label{sca15}
\eeq
where $\langle p \rangle$ is the mean momentum and
$\Delta p$ is the quantum mechanical uncertainty in
$p$ for this wave packet.  This wave packet needs to be evolved to
$-\tau$ using the free time evolution which adds a phase to (\ref{sca15}): 
\beq
\langle p \vert \psi_{0} (-\tau) \rangle =
{1 \over 2 \sqrt{\pi} \Delta p} e^{ - {(p-p_i)^2 \over 4 (\Delta p)^2}
  + i {p^2 \over 2 \mu} \tau}.
\label{sac16}
\eeq
In the discrete ``$p$'' representation this is replaced by 
\beq
\langle n \vert \psi_{0}(-\tau) \rangle = C e^{-{(\epsilon n - \langle p \rangle)^2 \over
4 (\Delta p)^2} + i {n^2 \epsilon^2 \over 2 \mu} \tau} .
\label{sac17}
\eeq
where $C$ is a normalization constant.
In the $x$ representation this becomes
\beq
\langle m \vert \psi_0 (-\tau) \rangle = {\epsilon \over \sqrt{2\pi}}
\sum_{n=-K}^K e^{i \epsilon^2 mn}
\langle n \vert \psi_{0}(-\tau) \rangle .
\label{sac18}
\eeq

To illustrate that this gives a good approximation to the continuum
results  $\langle p \rangle$,
$\langle x \rangle$, $\Delta p$ and $\Delta x$ were calculated starting
with $\langle p \rangle=2.5$, $\Delta p=.25$ and $K=300$ as input parameters in
(\ref{sac16}).  The results of the calculation
\begin{eqnarray*}
\mbox{mean}_{p-calc}&=2.500000\\
\mbox{mean}_{x-calc}&=0.000000\\
%\mbox{mean}_{x-calc}&=-2.51 \times 10^{-17}\\
\Delta_{p-calc}&=.3000000\\
\Delta_{x-calc}&= 1.666667
\end{eqnarray*}
are consistent with the input parameters,  the minimal uncertainty condition,
$\Delta p \Delta x = \frac{1}{2}$, 
and the continuum results.

With these states the sharp-momentum half-shell transition matrix elements are
\beq
\langle p_f \vert T(E_i) \vert p_i \rangle \approx
\langle \psi_{0f}(0) \vert V X^N  \vert \psi_{0i}(-\tau)\rangle.
\eeq
As a test
the discrete approximation
was applied to the problem of one-dimensional scattering of
particle of mass $m$ by a repulsive Gaussian potential of the form
\beq
V(x) = \lambda e^{-\alpha x^2}
\eeq
with $\lambda=.5$ and $\alpha =2.0$.
The potential is plotted in figure 1.  The particle's mass is taken to
be 1 in dimensionless units so the velocity can be identified with the
momentum.  The initial wave packet is a Gaussian with a delta
function normalization in momentum space with mean momentum $p=2.5$
and width $\Delta p= .25 $. It is pictured in figure 2.  The Fourier
transform of the initial wave packet is given in figure 3.  The oscillations
are because the momentum space wave packet has a non-zero mean momentum.  Given the
size of the potential and wave packets, the wave packet needs to move
about 18 units to the left in order to be out of the range of the
potential.  This suggest that for $v=p/m=2.5$ that $\tau=7$ should
be sufficient to move the wave packet out of the range of the
potential.  The resulting
free wave packet at $\tau=-7$ is shown in figure 4.  The scattered
wave function with $K=300$ ($M=601$) after $N=100$ time steps is shown
in figure 5,
and that
result multiplied by the potential is shown in figure 6.
Compared to the wave function in figure 3, the wave function in figure 5
includes the effects of the interaction.
Figure 6 illustrates the cutoff due to the short range potential; it
shows how only the part of the wave function inside the range of the
interaction contributes to the scattering operator.
Figure 7 compares the result of the off-shell Born approximation $\langle p
\vert V \vert \psi(0) \rangle$ to the calculation of the real and
imaginary parts of $\langle p \vert T \vert \psi(0) \rangle$ while
figure 8 compares $\langle p \vert T \vert \psi(0) \rangle$ to
$\langle p \vert T (p_0)\vert p_0 \rangle$ obtained by numerically
solving the Lippmann-Schwinger equation using the method \cite{Rubtsova}.

Figure 8 shows that the path integral computation with an initial wave
packet with a width of 1/10 of the momentum converges to the numerical
solution of the integral equation.  In unrelated time-dependent
scattering calculations \cite{Kopp:2011vv} a $\Delta p$ of about a tenth of
$p$ gave good approximations to sharp momentum matrix elements of the
transition operator for a wide range of momenta.

Unlike the solution of
the Lippmann Schwinger equation, in the path integral approach for
each energy it is necessary to determine minimal values of
$M$,$N$,$\tau$ and $\Delta p$ that are needed for convergence.
In practice there are a number of trade offs.  Making
the wave packets narrow in momentum increases the scattering volume
in the coordinate representation.  This in turn requires a larger $\tau$
to get out of the range of the potential.  If $\tau$ gets too large
the wave packet can move past $x_{max}=K\epsilon$ and will reappear
at $x_{min}=-K\epsilon$.  As $p$ gets large the oscillations in the
$x$ space wave function have higher frequencies, which requires
smaller time steps, while when $p$ gets small it is necessary to
make the wave packet width in momentum small enough so the
coordinate space tail of the wave function
gets out of the interaction volume.

The computations require storing the initial vector.  It is not
necessary to store the one-step transfer matrix - it can be computed
efficiently on the fly.  This is important for realistic calculations
since the vectors will be significantly larger in higher dimensions.
The hope is that in the future qubits can be used to represent large
vectors.

This one-dimensional example approximated half-shell
sharp-momentum transition matrix elements.  The on-shell values can be
used to extract other observables such as phase shifts and in the
one-dimensional case transmission and reflection coefficients.  This
formulation of the one-dimensional problem in terms of transition
matrix elements has the advantage that the method can be formally
be used in higher dimensions and to treat complex reactions or scattering
in quantum field theory.

The formulation of the discrete path integral used the
discrete Schwinger representation based on a single pair of
complementary operators where the complex one time step
probability is represented by a dense matrix.  An equivalent
representation in terms of qubits involves tensor products of
matrices (\ref{qb7}-\ref{qb8}) that act on single qubits, which may
have computational advantages.

One observation from these calculations that will have an impact on
future computations using quantum computers is that width of the wave
packets, total time for scattering, the scattering energy, the number
of time steps and the resolution of the discretization all have to be
considered together for an accurate and efficient calculation.  This is
because each one of these approximations generates its own source of
errors that impact the errors in the other approximations.

The general strategy discussed above can also principle be utilized to
formulate scattering quantum field theory.  The Haag-Ruelle
formulation of scattering
\cite{Haag:1958vt}\cite{Brenig:1959}\cite{Ruelle:1962}\cite{jost}
is the natural field-theoretic generalization of the quantum
mechanical treatment of time dependent scattering.  Like ordinary
quantum mechanical scattering it uses strong limits and needs one-body
(i.e. bound state) solutions to formulate the scattering asymptotic
conditions.

\begin{figure}
\begin{minipage}[t]{.45\linewidth}
\centering
\includegraphics[angle=000,scale=.4]{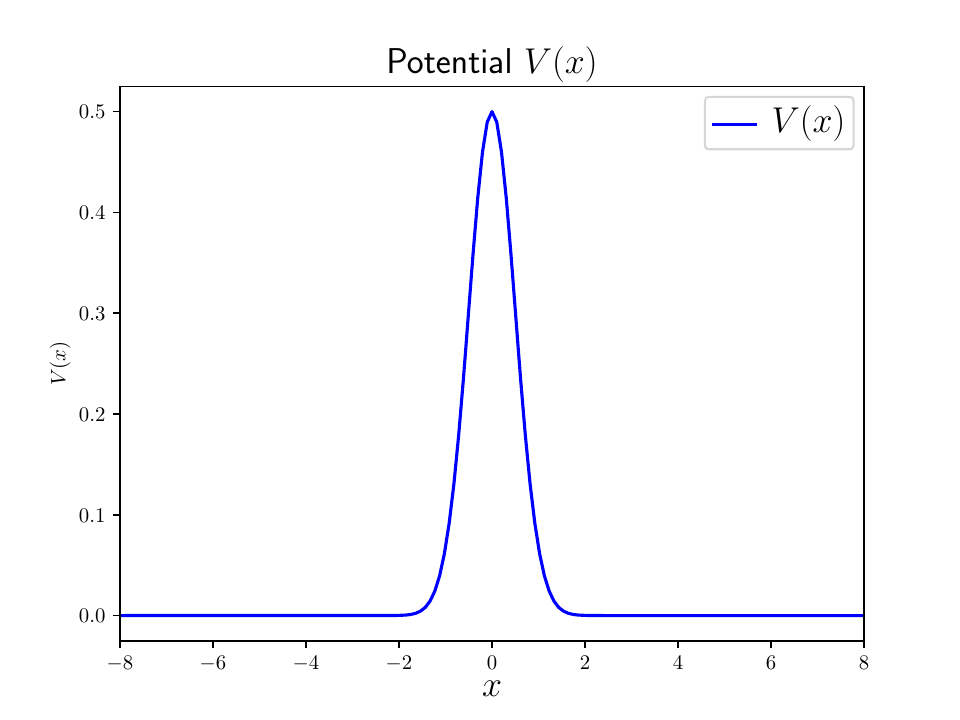}
\caption{\bf Potential}
\label{fig:1}
\end{minipage}
\begin{minipage}[t]{.45\linewidth}
\centering
\includegraphics[angle=000,scale=.4]{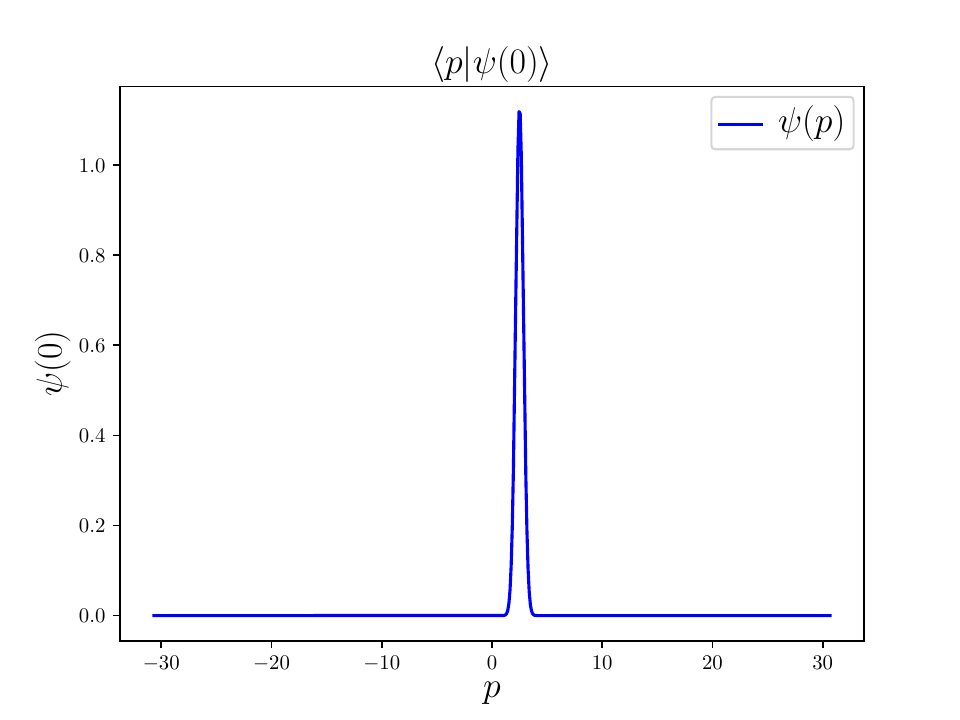}
\caption{\bf Momentum space initial Gaussian wave packet}
\label{fig:2}
\end{minipage}
\end{figure}

\begin{figure}
\begin{minipage}[t]{.45\linewidth}
\centering
\includegraphics[angle=000,scale=.4]{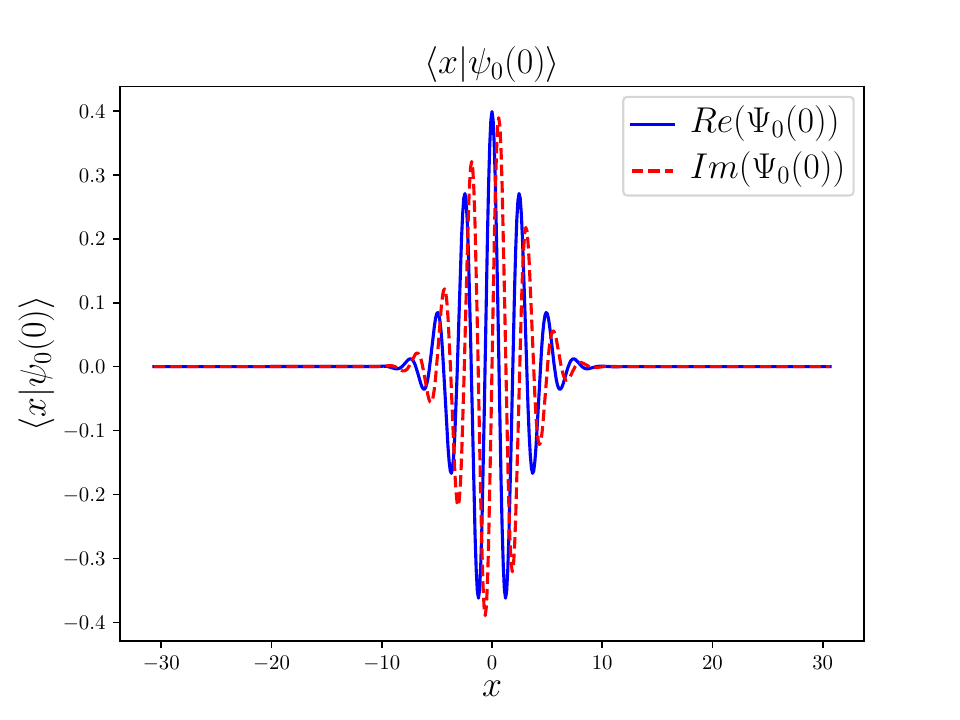}
\caption{\bf Coordinate space initial Gaussian wave packet}
\label{fig:3}
\end{minipage}
\begin{minipage}[t]{.45\linewidth}
\centering
\includegraphics[angle=000,scale=.4]{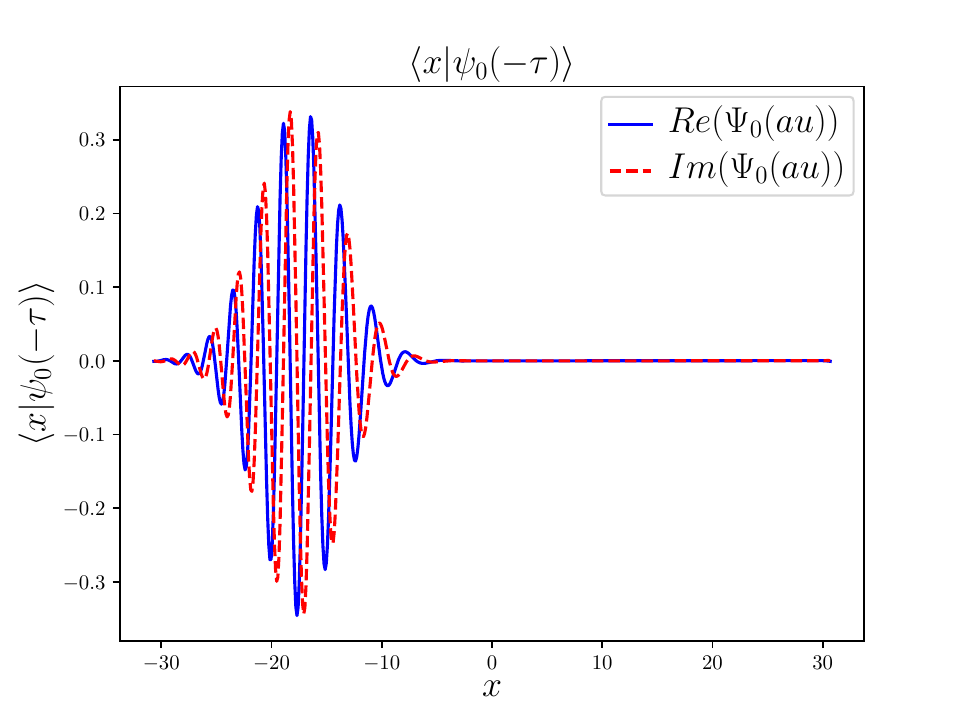}
\caption{\bf Free Gaussian wave packet at $\tau=-4$}
\label{fig:4}
\end{minipage}
\end{figure}

\begin{figure}
\begin{minipage}[t]{.45\linewidth}
\centering
\includegraphics[angle=000,scale=.4]{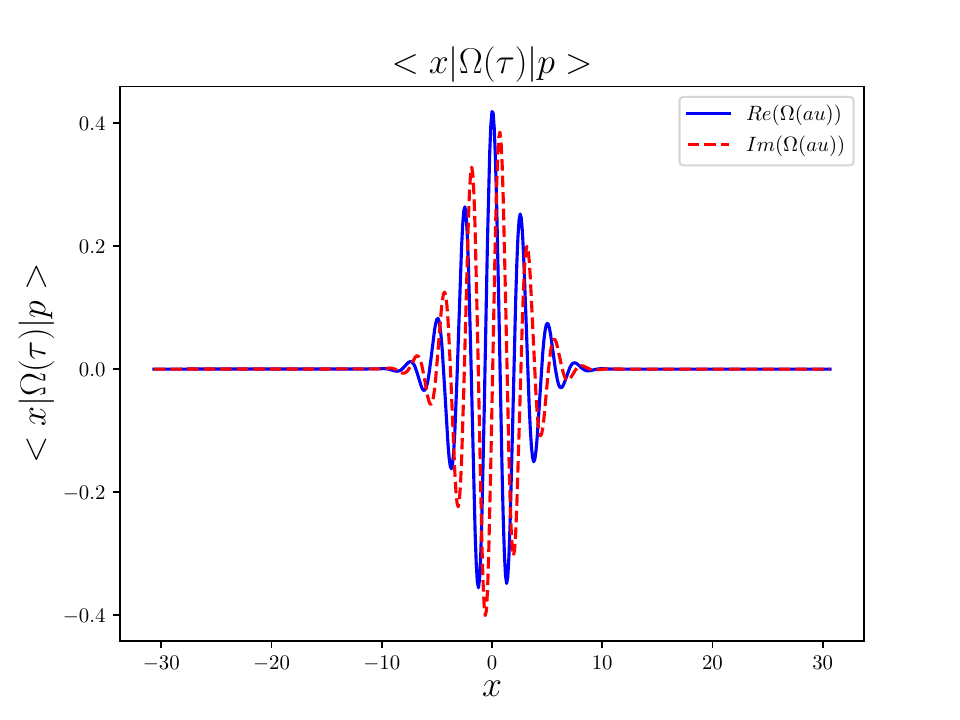}
\caption{\bf Initial scattering state at $t=0$ }
\label{fig:5}
\end{minipage}
\begin{minipage}[t]{.45\linewidth}
\centering
\includegraphics[angle=000,scale=.4]{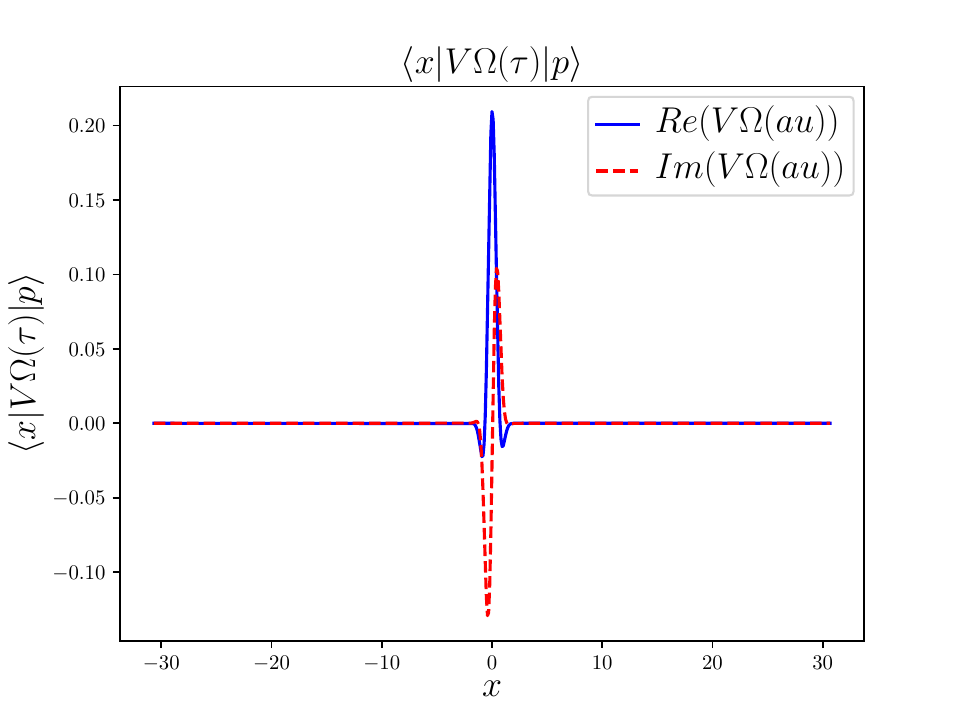}
\caption{\bf V $\times$ initial scattering state at $t=0$}
\label{fig:6}
\end{minipage}
\end{figure}

\begin{figure}
\begin{minipage}[t]{.45\linewidth}
\centering
\includegraphics[angle=000,scale=.4]{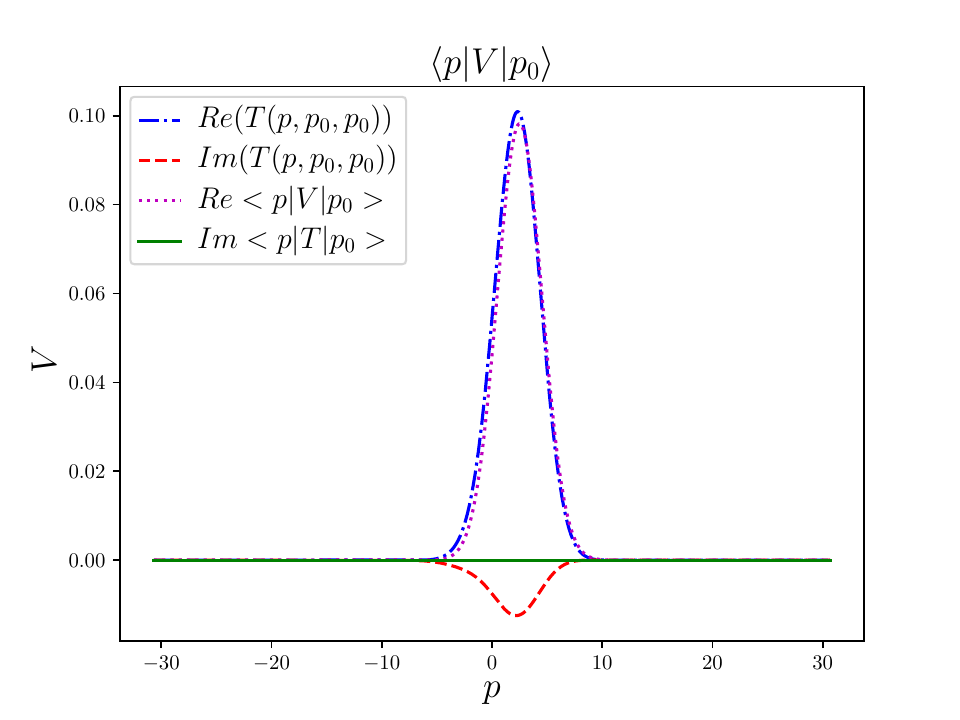}
\caption{\bf $\langle p \vert V \vert \psi_{0i}(0) \rangle$}
\label{fig:7}
\end{minipage}
\begin{minipage}[t]{.45\linewidth}
\centering
\includegraphics[angle=000,scale=.4]{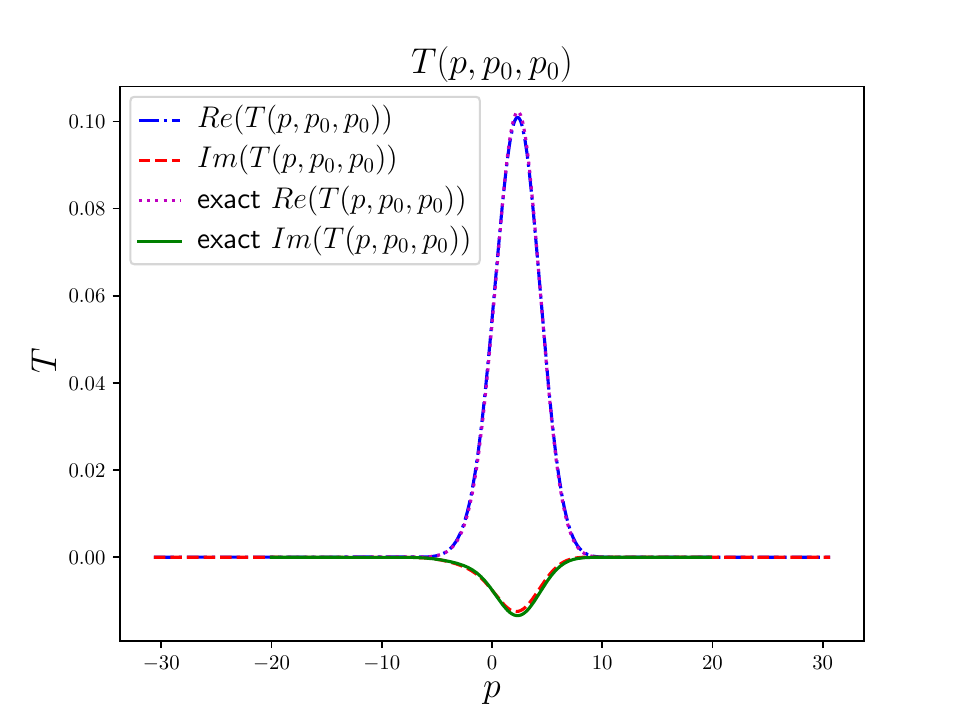}
\caption{\bf $\langle p \vert T \vert \psi_{0i}(0) \rangle$}
\label{fig:8}
\end{minipage}
\end{figure}

\bigskip\bigskip

\section{Discrete multi-resolution representation of quantum field theory}\label{qft}

One motivation for studying quantum computing in physics is that it
might provide a framework for a numerical treatment of problems in
strongly interacting quantum field theory.  Clearly this goal is a
long way off for realistic theories, but the state of quantum
computing is advancing rapidly.  Discrete formulations of field theory
naturally fit into the discrete framework discussed in this work and
should be relevant for future applications.  Discrete truncations to a
small number of degrees of freedom can be used as a laboratory to
explore how different field modes interact in a more realistic
truncation of the theory.

A numerical treatment of quantum field theory requires a truncation to
a system with a finite number of degrees of freedom.  For reactions
that take place in a finite space-time volume and involve a finite
energy it is natural to limit the number of degrees of freedom by
making volume and resolution truncations.  Degrees of freedom that are
outside of this volume or energetically inaccessible due to their
resolution are expected to be unimportant for the given reaction.
Daubechies wavelets
\cite{daubechies}\cite{jorgensen1}\cite{jorgensen2} and scaling
functions are a basis of square integrable functions and a natural
representation to perform both kinds of truncations.  The basis
consists of a complete orthonormal set of functions that have compact
support and a limited amount of smoothness.  They have the property
that in any small volume there are an infinite number of basis
functions supported entirely in that volume.  This means that they can
be used to construct ``local'' observables by smearing the fields with
basis functions.  All of the basis functions $\xi_n(x)$ are generated
from the solution of a linear renormalization group equation by
translations and dyadic scale transformations, which facilitates
computations.  The construction of the basis is discussed
in (\ref{app3}).
Because they are complete they can be used to {\it
exactly} expand the canonical fields
\[
\Phi (\mathbf{x},t) = \sum \Phi_n (t) \xi_n (\mathbf{x})
\qquad
\Pi (\mathbf{x},t) = \sum \Pi_n (t) \xi_n (\mathbf{x})
\]
where
\[
\Phi_n (t) := \int d\mathbf{x}\Phi (\mathbf{x},t)\xi_n (\mathbf{x}) 
\qquad
\Pi_n (t) := \int d\mathbf{x}\Pi (\mathbf{x},t)\xi_n (\mathbf{x}) 
\]
are discrete field operators.
If the fields satisfy canonical equal-time commutation relations
\beq
[\Phi (\mathbf{x},t), \Pi(\mathbf{y},t)] = i \delta (\mathbf{x}-\mathbf{y})
\eeq
then the discrete fields $\Phi_n$ and $\Pi_n$ will satisfy discrete
versions of the canonical equal time commutation relations
\cite{Bulut:2013bg} \cite{Polyzou:2017wnj}
\cite{Polyzou:2020ifj}: 
\[
[\Phi_m(t),\Pi_n(t)]=i \delta_{mn} 
\]
\[
[\Phi_m(t),\Phi_n(t)] =0
\]
\beq
[\Pi_m(t), \Pi_n(t)] =0 .
\label{w1}
\eeq
In terms of these degrees of freedom the Hamiltonian
for a $\phi^4$ theory has the form \cite{Bulut:2013bg} \cite{Polyzou:2017wnj}
\[
H= {1 \over 2}\sum_n \Pi_n\Pi_n + {m^2 \over 2}\sum_n \Phi_n\Phi_n +
\sum_{mn} D_{mn} \Phi_m \Phi_n +
\]
\beq
\lambda
\sum_{klmn} \Gamma_{klmn} \Phi_k\Phi_l\Phi_m\Phi_n
\label{wav2}
\eeq
where the sum are all infinite.  Since $H$ commutes with itself the
discrete fields in $H$ can be evaluated at $t=0$.

The equal time canonical commutation relations (\ref{w1}) imply that
the momentum modes generate translations of the field modes and the
field modes generate translations of the momentum modes.  This means
that the spectrum of both $\Phi_n$ and $\Pi_n$ is the real line.  In
the discrete approximation the amplitude of each field mode is
replaced by a large finite set of points that approach the continuum
following the construction in section \ref{cont}.  The cylinder sets
of paths for each mode are ``paths'' that take on the value of one of
these points at each time slice.  The total cylinder set is the
product of these sets over all retained $\Phi_n$ and $Pi_n$ modes.
This means that if $M$ $\Phi_n$ and $\Pi_n$ modes are retained, there
are $2M$ sequences of discrete values representing $2M$ cylinder sets of
paths.  In a truncation the discarded modes do not contribute to the
dynamics since their complex probabilities sum to 1, so it is only
necessary to determine the dynamics of the retained modes.

The matrices in (\ref{wav2}) are constants defined
by the integrals
\beq
D_{mn} = 
{1 \over 2} \int \pmb{\nabla} \xi_n (\mathbf{x})\cdot
\pmb{\nabla} \xi_m (\mathbf{x})d\mathbf{x} 
\label{wav3}
\eeq
\beq
\Gamma_{klmn}= \int
\xi_k (\mathbf{x}) \xi_l (\mathbf{x}) \xi_m (\mathbf{x}) \xi_n (\mathbf{x})
d\mathbf{x} 
\label{wav4}
\eeq
where
\[
\pmb{\nabla} =
\hat{\mathbf{x}}\frac{\partial}{\partial x}+
\hat{\mathbf{y}}\frac{\partial}{\partial y}+
\hat{\mathbf{z}}\frac{\partial}{\partial z}
\]
is the gradient operator.
For the wavelet basis
these constants vanish unless all of the functions appearing in the
integrals have a common support, which makes them almost local.  In
addition, because all of the functions in the integrand are related
to a single function by
translations and scale transformations, the
integrals can all be expressed as linear combinations of solutions of
some small linear systems of equations generated by the renormalization group
equation (see (\ref{w3})) \cite{Bulut:2013bg}
\cite{Polyzou:2020ifj}. Unlike a lattice truncation,
the wavelet representation
of the field theory is (formally) exact (before truncation).  The
basis functions regularize the fields so local products of fields that
appear in the Hamiltonian are replaced by infinite sums of
well-defined products of discrete field operators.  The basis functions are
differentiable, so there are no finite difference approximations.

Wavelet representations of quantum field theories have been discussed
by a number of authors
\cite{best:1994}
\cite{federbush:1995}
\cite{1995NuPhB.436..414H}
\cite{Battle:1999}
\cite{best:2000}
\cite{Ismail1:2003}
\cite{Ismail2:2003}
\cite{altaisky:2007}
\cite{albeverio:2009}
\cite{altaisky:2010}
\cite{altaisky:2010}
\cite{altaisky:2013}
\cite{Bulut:2013bg}
\cite{altaisky:2013b}
\cite{PhysRevA.92.032315}
\cite{PhysRevLett.116.140403}
\cite{altaisky:2016b}
\cite{altaisky:2016c}
\cite{altaisky:2016}
\cite{altaisky:2017}
\cite{Polyzou:2017wnj}
\cite{Neuberger2018}
\cite{Tomboulis1}
\cite{Polyzou:2020ifj}
\cite{Altaisky:2021hbq}
\cite{brennan3}.
What is relevant is that the Hamiltonian (\ref{wav2}) has the form
$\tilde{H} = \tilde{H}_1(p)+\tilde{H}_2(x)\to  \tilde{H}_1(\Pi)+\tilde{H}_2(\Phi)$,
except it involves an infinite number of degrees of
freedom.  It is diagonal and quadratic in the discrete momentum
operators and has a non-trivial (almost local) dependence on the
$\Phi_n$ operators.
%Because all of the basis functions are constructed
%from the fixed point $s(x)$ of the renormalization group equation
%(\ref{w3}) the constant quantities
%$D_{mn}$ and $\Gamma_{klmn}$ can be expressed in terms of a finite
%set of elementary integrals which are solutions of small linear systems.

The advantage of this basis is that it has natural volume and
resolution truncations.
%For reactions taking place in a finite volume
%with a finite energy a finite number of these degrees of
%freedom should provide a good approximation.
This reduces the problem
to a problem with a finite number of discrete degrees of freedom.  In
addition, the truncated Hamiltonian still has the form (\ref{wav2}),
except the sums are only over the retained discrete modes. As the volume and
resolution are increased (i.e as more modes are added) the parameters of
the theory have to be adjusted to keep the some physical observables
constant.  

The truncated problem is a finite number of degree of freedom
generalization of the one degree of freedom problem discussed in the
section \ref{scatt}.
%For a quantum field theory the vector
%representing the state of the field will be much larger than in the
%one degree of freedom scattering case.

In order to use this representation the constant coefficients 
$D_{mn}$ and $\Gamma_{n_1 \cdots n_k}$ that appear in the Hamiltonian (\ref{wav2})
need to be computed.
Using scale transformations (\ref{w4}) and the renormalization group
equation (\ref{w3}) they can all be expressed in terms of the integrals
\beq
d_{n} = 
\int {ds(x) \over dx} {ds(x-n) \over dx}dx \qquad -4\leq n \leq 4
\eeq
\beq
\gamma_{m,n,k}= \int s(x) s(x-m) s(x-n) s(x-k) dx \qquad -4 \leq m,n,k \leq 4 .
\eeq
These integrals for different values of $m,n,k$
are related to each other by finite linear equations
derived from the renormalization group equation (\ref{w3}) and the
scale fixing condition (\ref{w7}).  These linear systems can formally
be solved in terms of the coefficients $h_l$ (\ref{w5}).  The
coefficients $d_n$ are rational numbers and can be found in the
literature on wavelets \cite{beylkin1}.  To find the coefficients
$\gamma_{mnk}$ requires solving a system of $9^3$ linear equations.
%finding eigenvalues of a $9^3 \times 9^3$ matrix.
This
eliminates the need to evaluate fractal valued functions.
Alternatively the integrals $\gamma_{mnk}$ can be approximated by
noting that the renormalization group equation (\ref{w3}) and the
scale fixing condition (\ref{w7}) can be used to exactly calculate the
basis functions and their derivatives at all dyadic rational points.
Since the functions and their derivatives are continuous and the
dyadic rationals are dense this can be used to estimate these
quantities and integrals involving these quantities to any desired
accuracy.

In order to illustrate a path integral treatment of this system
consider a truncation of the theory in 1+1 dimensions 
where only 2 adjacent modes of the Hamiltonian (\ref{wav2}) are retained.
In this case the overlap coefficients that appear in the Hamiltonian
are related to $d_n$ and $\gamma_{lmn}$ by
\beq
\qquad D_{mn} = d_{n-m} \qquad \mbox{and} \qquad 
\Gamma_{klmn} = \gamma_{l-k,m-k,n-k}
\eeq
The coefficients that couple adjacent modes can be expressed in
terms of the following quantities
\begin{eqnarray*}
\Gamma_{0000}&= 0.9528539 \\
\Gamma_{0001}&= 0.0670946 \\
\Gamma_{0011}&= 0.0890895 \\
\Gamma_{0111}&=-0.1424536 \\
D_{00}&=295./56. \\
D_{01}&=-356./105. \\
D_{10}&=D_{01}\\
D_{11}&=D_{00}.   
\end{eqnarray*}
where the $\Gamma$ coefficients were computed by numerical integration
using the trapezoidal rule with the basis functions evaluated
at 256 dyadic points on their support.  Convergence was verified 
using 512 dyadic points. 

The truncated Hamiltonian in this case is
\[
H= {1 \over 2}\sum_{n=0}^1 \Pi_n\Pi_n + {m^2 \over 2}\sum_{n=0}^1 \Phi_n\Phi_n
+\sum_{m,n=0}^1 D_{mn} \Phi_m \Phi_n +
\]
\beq
\lambda
\sum_{k,l,m,n=0}^1 \Gamma_{klmn} \Phi_k\Phi_l\Phi_m\Phi_n
\label{wavxx}
\eeq
where
$\Gamma_{0000}=\Gamma_{1111}$,  $\Gamma_{0001}=\Gamma_{0010}=\Gamma_{0100}=
\Gamma_{1000}$, etc..
The path integral treatment of the field theory in the discrete representation
is a multi-dimensional generalization of the treatment for one degree of freedom
where each field mode represents an independent degree of freedom.

A general numerical treatment involves a truncation and
renormalization
followed by two approximations.  The truncation discards all but a
finite number, $F$, of discrete degrees of freedom.
\beq
H \to H_F .
\label{w22}
\eeq
Ideally physics at a given energy scale and in a given volume should
be dominated by a finite number of accessible degrees of freedom.  The
remaining degrees of freedom that are not expected to impact
calculations at that given scale and volume are discarded.  The
truncated theory is renormalized by adjusting the parameters of the
theory so a set of observables agree with experiment.  This gives the
parameters a dependence on the choice of retained degrees of freedom.
This is a truncation rather than an approximation. It assumes that no
additional parameters need to be introduced beyond what appears in the
truncated Hamiltonian and that there is a limit as the volume becomes
infinite and resolution becomes arbitrarily small.  This is followed
by two approximations.  The first approximation is to approximate the
unitary time evolution operator for the truncated theory using the
Trotter product formula with $N$ time slices:
\[
U_F(t) = e^{-i H_F t} =
\]
\beq
\lim_{N\to \infty} (e^{-i H_F(\Pi)\Delta t}e^{-iH_F(\Phi)\Delta t})^N 
\label{w23}
\eeq 
where $\Delta t = t/N$ and
\beq
H_F = H_F(\Pi) + H_F(\Phi)
\label{w24}
\eeq
with
\beq
H_F(\Pi) := {1 \over 2}\sum_n \Pi_n\Pi_n
\label{w24a}
\eeq
and
\[
H_F(\Phi):=
 {m^2 \over 2}\sum_n \Phi_n\Phi_n +
 \sum_{mn} D_{mn} \Phi_m \Phi_n +
 \]
 \beq
 \lambda
\sum_{klmn} \Gamma_{klmn} \Phi_k\Phi_l\Phi_m\Phi_n .
\label{w24b}
\eeq
This expresses $H_F$ as the sum of a part with only the $\Pi_n$
fields and another part with only the $\Phi_n$ fields.  Since the
discrete canonical pairs of field operators $\Phi_n$ and $\Pi_n$
satisfy canonical commutation relations they have a continuous
spectrum on the real line.  This is because each one of these
complementary operators generates translations in the other operator.
The last step is to approximate the continuous spectrum of the
discrete field operators $\Phi_n$ and $\Pi_n$ by a collection of
$M=2K+1$ closely spaced eigenvalues $\phi_n,\pi_n = n \epsilon$ where
$-K \leq n \leq K$ and $\epsilon^2 = 2 \pi/M$.  This is exactly what
was done in the scattering example, except in this case there are
$F$ degrees of freedom where $F$ is the number of retained discrete
field modes.  Unlike the truncation, both of these steps are
mathematical approximations.

Let $\langle \pmb{\phi} \vert \chi \rangle =
\chi(n_1\epsilon, \cdots , n_F \epsilon) $ be a localized function of the
amplitudes of the $F$ discrete field modes that represent an initial free wave
packet.

The goal is to use discrete path integrals to calculate the time evolution of
these coupled modes.  This gives a non-perturbative treatment of the
truncated problem.

For the field theory, before truncation, in the discrete
representation the path integral involves integrals over an infinite
number of modes.  The normalization of the complex probability $P_X$
is such that summing over all modes in the absence of interactions
gives 1.  This means that {\it the only modes that contribute
non-trivially to the time evolution are the retained modes}.  The
discrete approximation results in a sample space with a finite number
of discrete paths.

The Trotter approximation is 
\[
\langle
n_1,n_2,\cdots n_F \vert U_F(t) \vert \chi (0) \rangle =
\]
\beq
\lim_{N\to \infty} \langle
n_1,n_2,\cdots n_F \vert  (e^{-i H_F(\Pi)\Delta t}e^{-iH_F(\Phi)\Delta t})^N
\vert \chi (0) \rangle .
\label{w25}
\eeq
This can be evaluated by inserting complete sets of eigenstates of
the complementary fields 
between each of the operators.  The following abbreviations
are used for sums over intermediate states:
\beq
\int d\pmb{\phi} = \epsilon^F \sum_{n_1=-K}^K \cdots  \sum_{n_F=-K}^K,
\label{w26}
\eeq
for vectors representing a value of the eigenvalues of each of the $F$
independent $\phi$
field modes, 
\beq
\pmb{\phi}=(n_1 \epsilon, \cdots , n_F \epsilon) \qquad -K \leq n_i \leq K, 
\label{w27}
\eeq
for vectors representing the value of the eigenvalues of each of the $F$
independent $\pi$
field modes,
\beq
\pmb{\pi}=(n_1 \epsilon, \cdots , n_F \epsilon) \qquad -K \leq n_i \leq K 
\label{w28}
\eeq
and
\beq
\gamma = (\pmb{\phi}_0,\pmb{\phi}_1, \cdots, \pmb{\phi}_N)
\label{w29}
\eeq
for a ``path'' that ends at $\pmb{\phi}_0$ where
$\pmb{\phi}_j$ ($j>0$) 
represents values of each of the $\phi_n$ fields at
each of $N$ time steps.

The following definitions are generalizations of the definitions in
section \ref{path}:
\beq
P_X(\pmb{\phi}',\pmb{\phi},\Delta t) := \sum_{\mathbf{n}''}
\langle \pmb{\phi}' \vert \pmb{\pi}'' \rangle e^{-i \pmb{\pi}'' \cdot\pmb{\pi}'' \Delta t}
\epsilon^F\langle \pmb{\pi} \vert \pmb{\phi} \rangle \epsilon^F .
\label{w30}
\eeq
It follows from (\ref{f22}) that $P_X(\pmb{\phi}',\pmb{\phi},\Delta t)$ has the property
\beq
\sum_{\mathbf{n}} P_X(\pmb{\phi}',\pmb{\phi},\Delta t) =1
\label{w31}
\eeq
and
\[
P_{X}(\pmb{\phi}_f,\pmb{\phi}_N, \cdots \pmb{\phi}_1) :=
\]
\[
P_X(\pmb{\phi}_f,\pmb{\phi}_N,\Delta t)
P_X(\pmb{\phi}_N,\pmb{\phi}_{N-1},\Delta t)
\cdots
\]
\beq
P_X(\pmb{\phi}_3,\pmb{\phi}_2,\Delta t)
P_X(\pmb{\phi}_2,\pmb{\phi}_1,\Delta t)
\label{w32}
\eeq
also satisfies
\beq
\sum_{\gamma \in \Gamma}  
P_{X}(\pmb{\phi}_f,\pmb{\phi}_N, \cdots , \pmb{\phi}_1) =1 .
\label{w33}
\eeq
Equation (\ref{w32}) represents the complex probability of a given path, where at
each time slice each of the $F$ $\phi$'s has one of the $M$ allowed values
between $-K\epsilon$  and $K\epsilon$.  Removing the last factor
of $\epsilon^F$ and only summing over $\pmb{\phi}_N \cdots \pmb{\phi}_2$
gives the evolution due to free propagation
\beq
\langle
\pmb{\phi}_f \vert e^{-{i\over 2} \pmb{\Pi}\cdot \pmb{\Pi}t}
\vert \pmb{\phi}_1 \rangle =
\sum_{\pmb{n}_n \cdots \pmb{n}_1} 
P_X(\pmb{\phi}_f,\pmb{\phi}_N, \cdots , \pmb{\phi}_1)\epsilon^{-F} .
\label{w34}
\eeq
The full path integral including the effects of the interaction can
be expressed as the expectation of the following potential functional
of the path $\gamma$ 
\beq
W[\gamma] :=
e^{i\sum_{n} H_F (\pmb{\phi}_n)   \Delta t}
\label{w36}
\eeq
with respect to the complex probability
distribution (\ref{w33}),
where
$H_F (\pmb{\phi}_n)$ represents the value of the $\pmb{\phi}$-dependent part
of the Hamiltonian evaluated at the value of the path $\gamma$ at the
$n$-th time slice.

This gives the path integral approximation
\[
\langle
n_{1f},n_{2f},\cdots n_{Ff} \vert   U_F(t) \vert \chi (0) \rangle =
\]
\beq
\sum_{\gamma} 
P_X(\pmb{\phi}_f,\pmb{\phi}_N, \cdots \pmb{\phi}_1)W[\gamma]
\chi (\pmb{\phi}_1)
\label{w37}
\eeq
where $U_F(t)$ is the unitary time evolution operator (\ref{w23})
which again represents the path integral for fields as the expectation value
of
a potential functional with respect to a
complex probability distribution.  As in the one degree of freedom case
this can be exactly factored into a product of one-time step operators
\[
P_X(\pmb{\phi}_f,\pmb{\phi}_N, \cdots \pmb{\phi}_1)W[\gamma] =
\]
\[
X(\pmb{\phi}_f,\pmb{\phi}_N,\Delta t)
e^{iH_F (\pmb{\phi}_N)   \Delta t}
X(\pmb{\phi}_N,\pmb{\phi}_{N-1},\Delta t)
e^{iH_F (\pmb{\phi}_{N-1})   \Delta t}
\cdots
\]
\beq
X(\pmb{\phi}_3,\pmb{\phi}_2,\Delta t)
e^{iH_F (\pmb{\phi}_2)   \Delta t}  
X(\pmb{\phi}_2,\pmb{\phi}_1,\Delta t)
e^{iH_F (\pmb{\phi}_1)   \Delta t} .
\label{w38}
\eeq
This represents time evolution as the product of
large approximate transfer matrices.

Each stage of the calculation uses finite mathematics.   The use
of the finite Weyl representation exactly preserves unitary at each
level of approximation.  Both the $\pmb{\phi}$ and $\pmb{\pi}$
transfer matrices are unitary and can be expressed exactly in the
truncated model.  This means that the discrete Trotter approximation to
time evolution is exactly unitary.

Figures 9 and 10 show calculations of the initial
real and
imaginary parts of the two field modes.  In this case the
initial modes are real and taken to be Gaussians of the form
\beq
\langle \phi_1,\phi_2 \vert
\psi \rangle =
N e^{-\sum_{i=0}^1 (\phi_i - \langle \phi_i \rangle)^2/(4 \delta \phi_i^2)}
\eeq 
Figures 11 and  12 show the real and imaginary parts of 
the time $t=.5$ evolved amplitudes of the two
discrete modes with $M=41$ field values using 
$N=20$ Trotter steps.

Figures 13 and 14 show plots of the real and imaginary parts of $\phi_1$ when
$\phi_2=0$ at $T=0$ and $T=.5$. 

In the initial calculations the initial mean displacement and
uncertainty of each mode was taken to be .5. The initial state has no
imaginary part but one develops due to the non-zero displacement of
the initial state.  This truncation is too crude to contain any real
physics, however it illustrates the application of the discrete path integral
to fields.

A more drastic truncation of the discretization of the continuum could
be used to explore the dynamics of fields with a larger number of modes.

The wavelet representation for the Hamiltonian satisfies a functional
renormalization group equation that could be used to reduce the number
of amplitudes.  This relates infinite volume truncated Hamiltonians at different
resolutions using a canonical transformation along with 
mass, wave function, and coupling constant renormalizations: 
\[  
H_k (\Pi,\Phi,\mu,\lambda ) = 2^k H_0 (2^{-k}\Pi,2^{k}\Phi,2^{-2k}\mu,2^{-2k}\lambda]. 
\]

Realistic calculations require a large number of field modes.  Time-
dependent scattering calculations of the type discussed in section 9
also require volumes sufficiently large for the scattered fragments to
become stable particles.  These calculations cannot be performed on a
classical computer and will still be very challenging on a quantum
computer.

\bigskip
% calculations performed using daub.c
% graphs daub_r.gp, daub_i.gp unicode greek letters
%
\begin{figure}
\begin{minipage}[t]{.45\linewidth}
\centering
\includegraphics[angle=000,scale=.5]{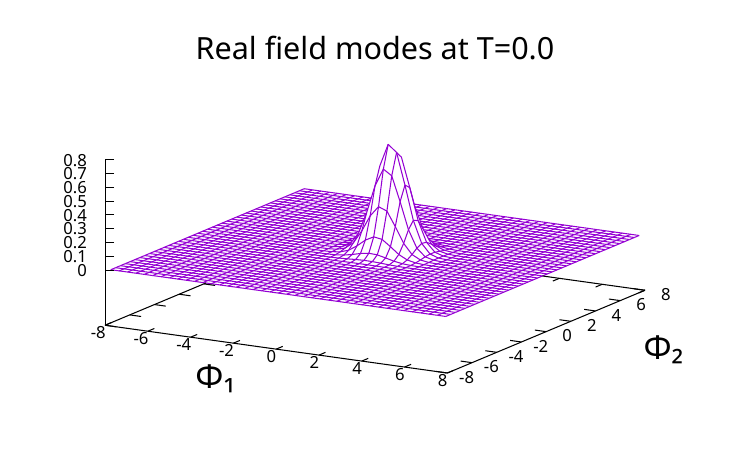}
\caption{\bf Two modes (real) at T=0.0}
\label{fig:9}
\end{minipage}
\begin{minipage}[t]{.45\linewidth}
\centering
\includegraphics[angle=000,scale=.5]{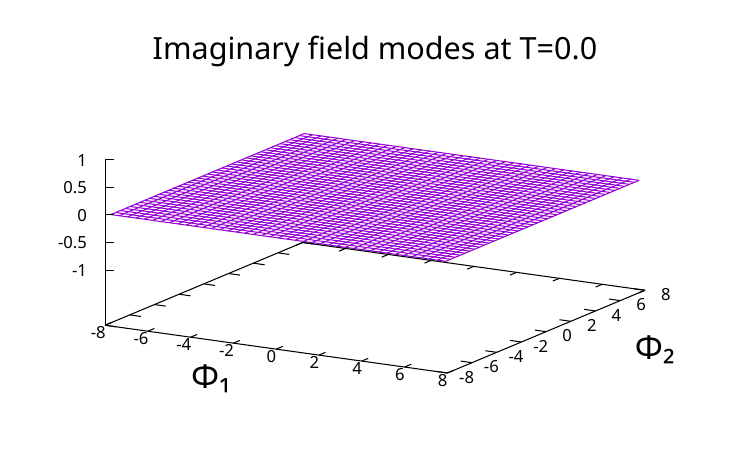}
\caption{\bf Two modes (imaginary) at T=0.0}
\label{fig:10}
\end{minipage}
\end{figure}

\begin{figure}
\begin{minipage}[t]{.45\linewidth}
\centering
\includegraphics[angle=000,scale=.5]{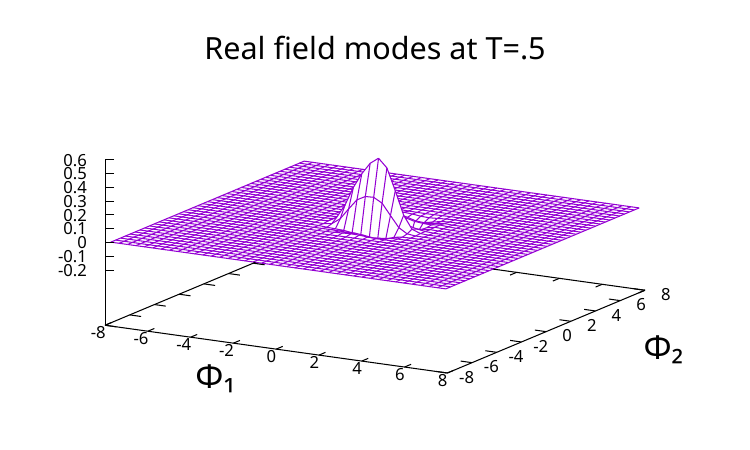}
\caption{\bf Two modes (real) at T=0.5}
\label{fig:11}
\end{minipage}
\begin{minipage}[t]{.45\linewidth}
\centering
\includegraphics[angle=000,scale=.5]{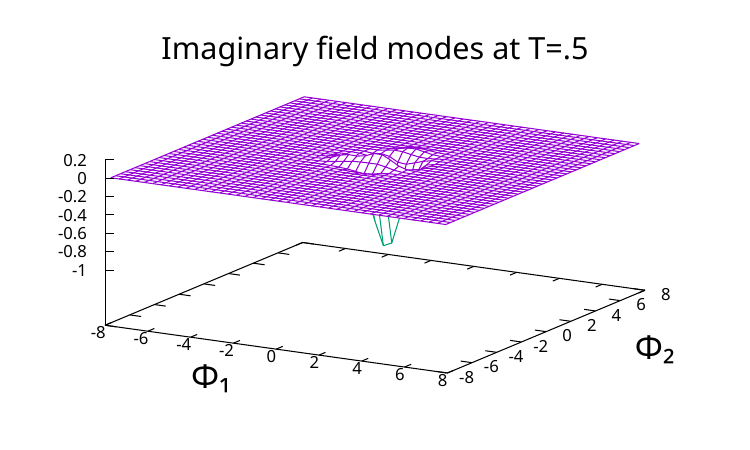}
\caption{\bf Two modes (imaginary) at T=0.5}
\label{fig:12}
\end{minipage}
\end{figure}

\begin{figure}
\begin{minipage}[t]{.45\linewidth}
\centering
\includegraphics[angle=000,scale=.4]{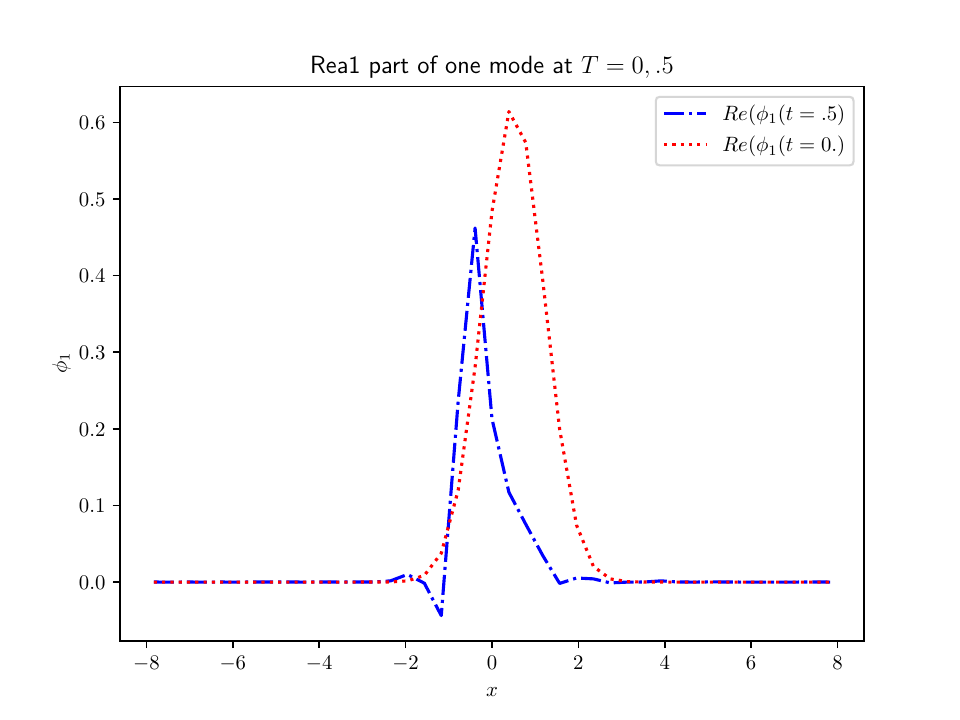}
\caption{\bf One mode (real) at T=0.0 and 0.5}
\label{fig:13}
\end{minipage}
\begin{minipage}[t]{.45\linewidth}
\centering
\includegraphics[angle=000,scale=.4]{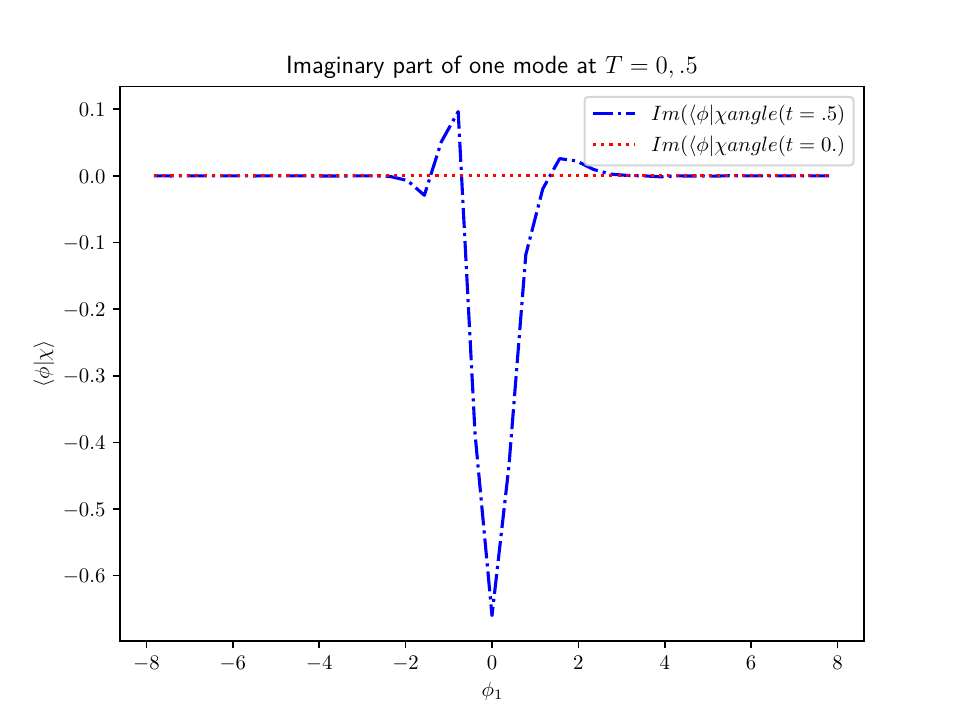}
\caption{\bf One mode (imaginary) at T=0.0 and 0.5}
\label{fig:14}
\end{minipage}
\end{figure}
  
\section{Summary and conclusion}\label{sum}

This paper discusses a complex probability interpretation of the
real-time Feynman path integral for systems of a finite number of degrees
of freedom.  In this interpretation the dynamics is expressed as the
expectation value of a functional of paths with respect to a complex
probability on a sample space of cylinder sets of ``paths''.  This
interpretation follows from the completeness relation and the Trotter
product formula.  The motivation is that the complex probability
interpretation also gives a rigorous reinterpretation of the real-time path
integral on infinite dimensional Hilbert spaces
\cite{Muldowney}\cite{Katya_1}\cite{Katya_2}, while the
treatment of real time evolution as a path integral is difficult to
interpret due to the absence of a countably additive positive measure
on the cylinder sets of paths.  For systems with a finite number of degrees
of freedom the complex probability interpretation is a natural
way to understand quantum dynamics,  and is closely related to
how the dynamics is treated in quantum computers.
%While the passage from
%imaginary time to real time is problematical for an interpretation of
%the path integral an integral, replacing the real-time integral by the
%expectation value of a functional on cylinder sets of paths with
%respect to a complex probability agrees with the Trotter limit.
%In the finite dimensional case the complex probability
%arises naturally by using the completeness relation at each time slice
%in the Trotter product.
In the ``phase space'' form, based on Schwinger's
\cite{schwinger} discrete Weyl representation, it can be applied to any
Hamiltonian matrix.  Infinite dimensional systems can be approximated as
limits of finite dimensional systems. 

The complex probability approach provides an alternative way of
thinking about path integrals.  It is difficult to interpret a
discrete-valued function of a continuous time variable as a path.  The
complex probability interpretation follows by observing that a
sequence of discrete eigenvalues at different time steps is a label
for a sequence of transition amplitudes from a state at the $n^{th}$
time step to a state at the $(n+1)^{th}$ time step.  In this sense a
cylinder set of ``paths'' can be thought of as an ordered sequence of
transition amplitudes between specific states at a sequence of time
steps.  Many ``paths'' contribute to a given cylinder set, since all
possible transition amplitudes are allowed at intermediate times
strictly between these successive time steps.  The complex probability
is the product of the probability amplitudes for a given sequence of
steps, while sum of all of these amplitudes gives the transition
amplitude for the transition from a given initial state to a final
state.  When this is summed over all final states the result is one
(the inner product of the initial state with itself).  With this
interpretation the Trotter product formula expresses the dynamics as
a random variable on this probability space.  The result of the
expectation value of this random variable with respect to the complex
probability is exactly, $\langle a_f \vert e^{-iHt} \vert a_i\rangle$,
which is an entire function of $t$ for a $D \times D$ dimensional
Hamiltonian $H$.

Computations of multi-particle scattering observables or quantum field
theory observables are candidates of problems that might be solved
with quantum computers.  These problems are formulated in infinite
dimensional Hilbert spaces.  For Hamiltonians that are sums of
non-commuting operators
%real-time evolution
the path integral representation of real-time evolution
%is naturally formulated, which
provides a means for
treating the non-commuting operators in the Hamiltonian without having
to explicitly diagonalize the Hamiltonian.

Quantum computers have a finite number of qubits, which means
that the dynamics must be approximated, where the approximate dynamics
is formulated on a finite-dimensional Hilbert space.  The
treatment of Hamiltonians with non-commuting operators is due to the
Trotter product formula.  When it is used in the finite dimensional
case paths are replaced by ordered sequences of discrete eigenvalues
of discrete Weyl operators.
%, which cannot be interpreted as paths.
Each finite sequence of discrete eigenvalues is associated with an ordered
product of transition amplitudes.

%When the Hamiltonian is set to zero the sum of all possible ordered
%products of transition amplitudes is 1 by completeness of the
%eigenstates of the unitary Weyl operators.  This is a finite sum
%(for a finite number of time steps) and justifies interpreting
%the ordered product of transition amplitudes as the complex probability
%for a sequence of eigenvalues of the discrete Weyl operators .

While this is not yet a model for quantum computing, any Hamiltonian
can be expressed as a polynomial in the discrete Weyl representation.
The Trotter product formula expresses time evolution as the
expectation value of a random variable on the space of sequences of
eigenvalues complementary pairs of discrete Weyl operators with
respect to the complex probability defined by the ordered product of
transition amplitudes.

A property of the discrete Weyl representation is that if the
dimension is factored into a product of prime numbers, the discrete
Weyl representation can be factored into a tensor product of
irreducible representations of prime dimension. If the dimension is a
power of 2 (i.e. $2^N$) then the Hilbert space can be represented by
$N$ qubits, and the Weyl operators acting on each qubit can be
represented by pairs of Pauli matrices.

In this case the $2^N$ Weyl operators represent a set of $2^N$ qubit
gates, which by construction are irreducible.  Products of
these gates can be used to build two qubit gates.  A ``quantum gate''
in the original Weyl representation is replaced by a product of gates
in the qubit representation.  While mathematically equivalent, the
qubit representation is the natural representation for quantum devices.
In both cases the complex probability interpretation remains
unchanged.

An important property of the complex probability for finite
dimensional Hilbert spaces is that it exactly factors into products of
conditional probabilities at each time step.  This factorization also
applies to the dynamics which modifies the phase of the transition
amplitude at each time step (see \ref{cpi18}).  The phase modification
at each time step is represented by a quantum circuit. The factorization
is also an important computational simplification.

%It was shown that the discrete representation has an equivalent qubit
%representation in terms of an irreducible set of elementary qubit
%gates.
In the qubit representation the discrete intermediate states are
labeled by the states of all qubits at each time step, while quantum
circuits define the random variable associated with time evolution.
The qubit representation is natural for quantum computations since the
complementary operators, $U_i$ and $V_i$ act on single qubits.  These
operators are the irreducible building blocks in this representation.
This representation is efficient for treating local operators that can
be expressed in terms of a small number of qubit operators.  The two
examples with continuous spectra illustrate how the direct approach
can be used to investigate convergence with respect to the number of
time steps and resolution before passing to a qubit representation.

Hamiltonians for many realistic systems act on infinite dimensional
Hilbert spaces.  In this work these systems are treated as limiting cases
of large finite dimensional systems rather than finite dimensional
truncations of infinite dimensional systems.

Discrete approximations of infinite dimensional systems are discussed
in section \ref{cont}.  These approximations were used to demonstrate
the application of the discrete formulation of the path integral to
potential scattering and quantum field theory.  The path integral
approach is not the most efficient way to perform scattering
calculations, but the simple calculation presented in section
{\ref{scatt}, based time-dependent scattering theory, illustrates
some of the issues that need to be considered in future realistic
treatments of scattering using quantum computers.

While the goal of scattering calculations is to compute sharp-momentum
scattering amplitudes, time-dependent scattering requires strong
limits, which means that the sharp-momentum states must be replaced by
wave packets.  In the discrete case there are normalizable
sharp-momentum states, but they are not acceptable for scattering
because the corresponding position states are completely delocalized,
so the scattered particle can never get out of the range of the
potential.  In addition efficient calculations require paying careful
attention to the range of the interaction and both the width the both
the momentum and position wave packets, which cannot be controlled
independently.

In contrast to scattering problems, path integrals are one of the more
direct methods to solve interacting field theories.  The illustrated
application to field theory in section \ref{qft} uses a basis of
wavelets to replace the fields by an infinite collection of almost
local operators.  Volume and resolution truncations are used to
replace the field by a finite number of discrete modes. The
construction of the multi-resolution wavelet basis is discussed in
\ref{app3}. The calculations presented in section \ref{qft} use a
severe truncation of a $\phi(x)^4$ field theory to two modes, but the
calculation is completely non-perturbative.  Realistic calculations
are still a long way off.

%\ack{{This work supported by the U.S. Department of Energy,
%  Office of Science, Grant \#DE-SC16457,
%The author would like to thank William Hester for pointing out some errors
%in the original version of this manuscript.
%}

\appendix
  
\section{Schwinger's discrete Weyl algebra}
\label{app2}

This section reviews Schwinger's \cite{schwinger} method of
constructing an irreducible algebra of complementary unitary operators
for quantum systems of a finite number of degrees of freedom.  This
construction generates a finite degree of freedom version of the Weyl
(exponential) form of the canonical commutations relations.  This
algebra can be used to build discrete models of any finite quantum system.
%This construction is essentially the same as the treatment of the
%quantum Fourier transform discussed in \cite{nielsen} and elsewhere.

Let $X$ be a quantum observable with $M$ orthonormal eigenvectors
$\vert m \rangle$ associated with measurement outcomes $x_m$.  $X$
acts on a $M$ dimensional complex Hilbert space ${\cal H}$.
The eigenvectors of $X$ are a basis on ${\cal H}$:
\beq
X \vert m \rangle =x_m \vert m \rangle  \qquad m=1,\cdots, M .
\label{s1}
\eeq
Schwinger defines a unitary operator $U$ on ${\cal H}$ that cyclically shifts the
eigenvectors of $X$:
\beq
U\vert m \rangle = \vert {m+1} \rangle \qquad m<M
\qquad
U\vert M \rangle = \vert 1 \rangle .
\label{s2}
\eeq
The labels $m$ on the eigenvectors are treated
as integers mod $M$ so 0 is identified with $M$, $1$ with $M+1$ etc..
%$U$ defined by (\ref{s2}) is unitary since
%\beq
%UU^{\dagger}= \sum_{m=1}^M U\vert m \rangle \langle m \vert U^{\dagger}=
%\sum_{m=1}^{M-1} \vert {m+1} \rangle \langle {m+1} \vert 
%+ \vert {1} \rangle \langle {1} \vert =
%\sum_{m'=1}^M \vert {m'} \rangle \langle {m'} \vert = I .
%\label{s3}
%\eeq
Since $M$ applications of $U$ leaves all $M$
basis vectors, $\vert m \rangle$,  unchanged, it follows that $U^M=I$.
Since $U^k\vert m \rangle$ are independent for all $k<M$, there are no
lower degree polynomials in $U$ that vanish, so 
$P(\lambda) = \lambda^M-1 =0$ is the characteristic polynomial of $U$.
The eigenvalues $\lambda$ 
of $U$ are the $M$ roots of $1$:
\beq
\lambda = u_m = e^{2\pi  m i\over M}
\label{s3}
\eeq
with orthonormal eigenvectors $\vert u_m \rangle$:
\beq
U \vert u_m \rangle = u_m \vert u_m \rangle
\label{s4}
\eeq
\beq
\langle u_m \vert u_n \rangle = \delta_{mn}.
\label{s5}
\eeq
The normalization (\ref{s5}) does not fix the phase of the $\vert u_n \rangle$ 
which will be chosen later.  Both $U^M=I$  and $u_n^M=1$
imply that
\[
0=
(U^M-I ) = {1 \over u_n^M}  (U^M-I  )=
({U\over u_n})^M-I =
\]
\beq
%\prod_{m=1}^M ({U\over u_n}- {u_m \over u_n}) =
\Big(\frac{U}{u_n}-I \Big)\Big(I +\frac{U}{u_n} + \Big(\frac{U}{u_n}\Big)^2 +\cdots + \Big(\frac{U}{u_n}\Big)^{M-1} \Big).
\label{s6}
\eeq
Since this expression is identically zero and $({u_m\over u_n}-1)\not=0$
for $m\not= n$ it follows that
\beq
I +{U\over u_n} + \Big({U\over u_n}\Big)^2 +\cdots \Big({U\over u_n}\Big)^{M-1}
= c \vert u_n \rangle \langle u_n \vert 
\label{s7}
\eeq
for some constant $c$.
Applying (\ref{s7}) to $\vert u_n \rangle$ implies that the constant $c=M$.
This results in an expression for the projection operator on each eigenstate
of $U$ as a degree $M-1$ polynomial in $U$
\beq
%\boxed{
\vert u_n \rangle \langle u_n \vert =
{1 \over M} \sum_{m=1}^M \Big({U\over u_n}\Big)^m =
{1 \over M} \sum_{m=0}^{M-1} \Big({U\over u_n}\Big)^m .
%}
\label{s8}
\eeq
Using (\ref{s8}) it follows that
\[
\langle k \vert u_n \rangle \langle u_n \vert k \rangle  =
\]
\[
{1 \over M} \sum_{m=0}^{M-1} \langle k \vert \Big({U\over u_n}\Big)^m \vert k \rangle 
=
\]
\[
{1 \over M} \sum_{m=0}^{M-1} \Big({1\over u_n}\Big)^m\langle k \vert {k+m} \rangle
\]
\beq
= {1 \over M} .
\label{s9}
\eeq
This means that for any $k$ and $n$ that
\beq
\vert \langle k \vert u_n \rangle \vert = {1 \over \sqrt{M}} .
\label{s10}
\eeq
The interpretation is that if the system is prepared in any eigenstate
of $U$ and $X$ is subsequently measured, then the probability of
measuring any of the eigenvalues of $X$ is the same, (1/M).  This means that
all of the information about the identity of the initial eigenstate of
$U$ is lost after measuring $X$.  This is the condition for the
observables $X$ and $U$ to be complementary.

It is convenient to choose the phase of each $\vert u_n \rangle$
by
\beq
\langle M \vert u_n \rangle = \langle u_n \vert M \rangle
= {1 \over \sqrt{M}} .
\label{s11}
\eeq
It follows from (\ref{s11}) and (\ref{s8})  that
\[
\langle k \vert u_n \rangle \langle u_n \vert M \rangle =
\langle k \vert u_n \rangle {1 \over \sqrt{M}} =
\]
\[
{1 \over M} \langle k \vert \sum_{m=1}^M u_n^{-m} \vert m \rangle =
{1 \over M} u_n^{-k} =
\]
\beq
{1 \over M} e^{-2\pi i nk/M}.
\label{s12}
\eeq
Multiplying (\ref{s12}) by $ \sqrt{M}$ shows that the phase convention
(\ref{s11}) fixes the inner product $\langle k \vert u_n \rangle$
for all $k\not=M$:
\beq
\langle k \vert u_n \rangle =
{1 \over \sqrt{M}} e^{-i\frac{2\pi  nk}{M}}.
\label{s13}
\eeq
Schwinger defines another unitary operator, $V$, that shifts the eigenvectors of
$U$ cyclically, but in the opposite direction
\beq
V \vert u_n \rangle = \vert u_{n-1} \rangle, \qquad n \not=1,
\qquad
V \vert u_1 \rangle = \vert u_{M} \rangle . 
\label{s14}
\eeq
The same method, with $U$ replaced by $V$, gives
\beq
V^M=I
\label{s15}
\eeq
\beq
V \vert v_m \rangle = v_m \vert v_m \rangle
\qquad v_m = e^{2 \pi i m \over M}
\label{s16}
\eeq
\beq
\vert v_n \rangle \langle v_n \vert
=
{1 \over M} \sum_{m=0}^{M-1} \Big({V \over v_n}\Big)^m =
{1 \over M} \sum_{m=1}^{M} \Big({V \over v_n}\Big)^m
\label{s17}
\eeq
and for unit normalized $\vert v_n \rangle$
\beq
\vert \langle u_k \vert v_n \rangle\vert = {1 \over \sqrt{M}}.
\label{s18}
\eeq
It is convenient to choose the  
phase of  $\vert v_n \rangle$ by the condition 
\beq
\langle u_M \vert v_n \rangle = {1 \over\sqrt{M}}. 
\label{s19}
\eeq
then (\ref{s17}) and the orthonormality of the
$\vert u_k \rangle$'s give
\[
\langle u_M \vert v_n \rangle \langle v_n \vert u_k \rangle =
\langle v_n \vert u_k \rangle {1 \over \sqrt{M}} =
\]
\beq
{1 \over M} \sum_{m=0}^{M-1} v_n^{-m} \langle u_m\vert u_k \rangle =
{1 \over M} v_n^{-k}. 
\label{s20}
\eeq
Multiplying (\ref{s20}) by $\sqrt{M}$ gives
\beq
%\boxed{
\langle v_k \vert u_n \rangle = {1 \over \sqrt{M}} e^{-i\frac{2\pi  nk}{M}}.
%}
\label{s21}
\eeq
Comparing (\ref{s13}) and (\ref{s21}) it follows that
\[
\vert v_k \rangle =\sum_{m=0}^{M-1} \vert u_m \rangle \langle u_m \vert v_k \rangle=
\]
\[
\sum_{m=0}^{M-1} \vert u_m \rangle {e^{i\frac{2 \pi  mk}{M}}\over \sqrt{M}} = 
\]
\beq
\sum_{m=0}^{M-1} \vert u_m \rangle \langle u_m \vert k \rangle =
\vert k \rangle ,
\label{s22}
\eeq
so, because of the phase choices,  the operators $X$ and $V$ have the same eigenvectors.

The spectral expansion of $V$, the identification (\ref{s22}) of
$\vert v_k \rangle$ with $\vert k\rangle$ and the definition
(\ref{s2}) of $U$ imply that the 
unitary operators $U$ and $V$ satisfy
\[
UV = U \sum_{m=0}^{M-1} \vert v_m \rangle e^{i\frac{2 \pi m}{M}}
\langle v_m \vert =
\]
\[
\sum_{m=0}^{M-1} \vert v_{m+1} \rangle e^{i\frac{2 \pi m}{M}}
\langle v_m \vert =
\sum_{m=0}^{M-1} \vert v_{m+1} \rangle e^{i\frac{2 \pi m}{M}}
\langle v_{m+1} \vert U=
\]
\[
\sum_{m=0}^{M-1} \vert v_{m+1} \rangle e^{i2 \pi (m+1) \over M}
e^{-{2 \pi i \over M}}
\langle v_{m+1} \vert U=
\]
\beq
e^{-i\frac{2 \pi }{M}} VU .
\label{s23}
\eeq
%or
%\beq
%\boxed{
%UV = VU e^{-{2\pi i \over M}}.
%%}
%\label{s24}
%\eeq
$U$ and $V$ generate a complete set of operators in the sense that 
that any
operator on the Hilbert space can be expressed as a degree $(M-1) \times (M-1)$ polynomial
these two operators.  Note that $m-k$ applications of (\ref{s2})
give $U^{m-k} \vert v_k\rangle = \vert v_m\rangle$.
When $m-k$ is negative the operator is either the $(k-m)^{th}$ power of the
inverse (adjoint) of $U$ or equivalently it can be replaced by 
$M+m-k$ since $U^M=I$.
This also applies to the equations that follow.
Using this
with (\ref{s17}) gives
\[
\vert v_m \rangle \langle v_k \vert   =
U^{m-k} \vert v_k \rangle \langle v_k \vert   =
\]
\beq
{1 \over M} \sum_{n=0}^{M-1} e^{-i\frac{2 \pi  nk}{M}} U^{m-k} V^n .
\label{s25}
\eeq
The order of the $U$ and $V$ operators can be changed using
multiple applications of (\ref{s23}):
\[
U^m V^n = U^{m-1}V^n U e^{-i\frac{2n \pi}{M}} = 
U^{m-2}V^n U^2 e^{-{4n \pi i \over M}} =
\]
\beq
\cdots =
V^n U^m e^{-i\frac{2mn \pi}{M}}.
\label{s25a}
\eeq
Using (\ref{s25a}) in (\ref{s25}) gives
\[
\vert v_m \rangle \langle v_k \vert
\]
\[
{1 \over M} \sum_{n=0}^{M-1} e^{-i\frac{2 \pi nk}{M}} V^n U^{m-k}
e^{-i\frac{2(m-k)n \pi}{ M}} =
\]
\beq
{1 \over M} \sum_{n=0}^{M-1} e^{-i\frac{2 \pi mk}{M}} V^n U^{m-k} .
\eeq
A general operator $O$ can be expressed in terms of its matrix elements
in a basis
\[
O = \sum_{m,k=0}^{M-1} \vert v_m \rangle \langle v_m \vert O \vert v_k \rangle
\langle v_k \vert =
\]
\[
\sum_{m,k=0}^{M-1} \langle v_m \vert O \vert v_k \rangle
\vert v_m \rangle  \langle v_k \vert =
\]
\[
{1 \over M} \sum_{m,n,k=0}^{M-1} e^{-i\frac{2 \pi  nk}{M}} \langle v_m \vert O \vert v_k \rangle
U^{m-k} V^n =
\]
\beq
{1 \over M} \sum_{m,n,k=0}^{M-1} e^{-i\frac{2 \pi mn}{M}}\langle v_m \vert O \vert v_k \rangle V^n U^{m-k}.
\label{s26}
\eeq
These equations have the form
\beq
%\boxed{
O = \sum_{m,n=0}^{M-1} c_{mn} U^m V^n = \sum_{m,n=0}^{M-1} d_{mn} V^m U^n 
%}
\label{s27}
\eeq
which is the discrete Weyl representation of $O$. 
If $O$ commutes with $U$ then, using (\ref{s25a}), 
\[
0=  \sum_{mn=0}^{M-1} c_{mn} [U^m V^n,U] =
\]
\beq
\sum_{mn=0}^{M-1} c_{mn} U^{m+1} V^n
(e^{i\frac{2\pi  n}{M}}-1)
\label{s28}
\eeq
which requires $n=M$ or $0$. This means $O$ is independent of $V$.
Similarly if $O$ commutes with $V$ it must be independent of $U$.  It
follows that any operator that commutes with both $U$ and $V$ is a
constant multiple of the identity. This means that the operators $U$
and $V$ are an irreducible set of unitary operators.

\section{Hamiltonians that are sums of non commuting operators}
\label{app1}
The formalism discussed in (\ref{sec:findim}) can be applied to any
Hamiltonian on the $M$-dimensional Hilbert space.  If the Hamiltonian
is a sum of the form $\tilde{H}(p,x)=\tilde{H}_1(p)+\tilde{H}_2(x)$
then it is possible to sum over the intermediate ``momentum''
variables, eliminating $\tilde{H}_1(p)$, resulting in a complex
probability on a space of ``paths'' in the coordinate variable, $X$.
As in the phase space case, these ``paths'' have discrete values, so
they are generally nowhere continuous.

In this case
\[
\sum_n \langle k \vert \bar{n}\rangle \langle \bar{n}  \vert H \vert {m} \rangle =
\]
\[
\sum_n \langle k \vert \bar{n}\rangle \langle \bar{n} \vert m \rangle (H_1(p_n)+H_2(x_m))=
\]
\beq
\sum_n \langle k \vert \bar{n}\rangle \langle \bar{n} \vert m \rangle \tilde{H}_{nm}
\label{f19b}
\eeq
where $\tilde{H}_{nm}$ is just the ``classical'' Hamiltonian as a function
of the eigenvalues.

A new complex probability can be defined by 
\[
P_X(k_f; k_N,\cdots ,k_1)
:=
\]
\beq
e^{i \tilde{H}_1(p_0) t}
\sum_{n_1 \cdots n_N}
P(k_f; n_N,k_N,\cdots n_1,k_1)
e^{-i \sum \tilde{H}_1(p_{n_m}) \Delta t} . 
\label{f20}
\eeq
To show that this is normalized like a complex probability,
interchange the order of the sum over the intermediate
$k$ indices and $n$ indices (both sums are finite).  The
$k_2 \cdots k_N$ sums are just expressions for the identity.
After eliminating $k_2 \cdots k_N$ what remains is a product of
Kronecker delta functions in the $n_i$ variables.
Since the operator
$H_1$ is a multiplication operator in the $n$ variables, 
all but one of the $n$ sums can be performed giving:
\[
\sum_{n_1 \cdots n_N} \sum_{k_1 \cdots k_N} 
P(k_f; n_N,k_N,\cdots n_1,k_1)
e^{-i \sum \tilde{H}_1(p_{n_m}) \Delta t} =
\]
\[
\sum_{n=0}^{M-1}\sum_{k_1=0}^{M-1}\langle k_f \vert \bar{n} \rangle e^{-i N \tilde{H}_1(p_{n}) \Delta t}
\langle \bar{n} \vert k_1 \rangle .
\]
The $k_1$ sum can be evaluated using  
\[
\sum_{n=0}^{M-1}\sum_{k_1=0}^{M-1} \langle k_f \vert \bar{n} \rangle
e^{-i \tilde{H}_1(p_n) t}\langle \bar{n} \vert k_1 \rangle
=
\]
%\sum_{mn=0}^M \langle n_f \vert n \rangle =1 
%\]
%sum_{m,n=0}^M \langle n_f \vert \bar{m} \rangle \langle \bar{m} \vert n \rangle
\[
{1 \over M} \sum_{n=0}^{M-1} e^{-i\frac{ 2 \pi  k_f n}{M}} \sum_{k_1=0}^{M-1}
e^{i 2 \pi  k_1 n/M} e^{-i \tilde{H}_1(p_n) t} =
\]
\[
{1 \over M} \sum_{n=0}^{M-1}  e^{-i\frac{ 2 \pi  k_f n}{M}} e^{-i \tilde{H}_1(p_n) t}
\times
\]
\[
\left \{
\begin{array}{cc}
  (1 - e^{i 2 \pi  M n/M})=0 & 1<n<M-1\\
  M & n=0
\end{array}
\right .
\]
\beq
= e^{-i H_1(p_0) t}.
\label{f21}
\eeq
%xxxxxxxxxxxxxxxxxxxxxxxxxxxxxxxxxxxxxxxxxxxxxxxxxxxxxxxxxxxxxxxxxxxx
%Using this in (\ref{f20}), summing over the right most $k$ sets the next $n$ to% 0, then the process repeats, moving right to left resulting in 
%xxxxxxxxxxxxxxxxxxxxxxxxxxxxxxxxxxxxxxxxxxxxxxxxxxxxxxxxxxxxxxxxxxxxxxxxxx
Including the factor $e^{i H_1(p_0) t}$ gives 
\[
\sum_{k_1 \cdots k_N} P_X(k_f;,k_N,\cdots ,k_1)=
\]
\beq
e^{i \tilde{H}_1(p_0) t}
\sum_{n=0}^{M-1}\sum_{k_1=0}^{M-1}
P(k_f; n_N,k_N,\cdots n_1,k_1)
e^{-i \sum_m \tilde{H}_1(p_{m}) \Delta t} =1.
\label{f22}
\eeq
The expression for the evolution operator becomes
\[
\langle k_f \vert e^{-iHt} \vert k_i \rangle =
\]
\beq
\sum_{k_1 \cdots k_N} P_X(k_f;,k_N,\cdots ,k_1)
e^{-i \sum_m \tilde{H}_{2}( x_{n_m})  \Delta t}\delta_{k_1k_i}
\label{f23}
\eeq
which is expressed as the expectation value of the functional
$e^{-i \sum_m \tilde{H}_{2}( x_{n_m})  \Delta t}$ on cylinder sets of discrete paths.
The finite sums over the discrete intermediate ``momentum'' variables
replace the Gaussian Fresnel integrals in the continuum case.  

In this case the complex probability over cylinder sets of paths in
$x$ also factors into products of one-step probabilities with 
\beq
P_{X}(k_{n+1};k_n)=
\sum_{m}
P(k_{n+1}; m,k_n)
e^{-i (\tilde{H}_1(p_{m})-\tilde{H}_1(p_{0})) \Delta t} 
\label{f24}
\eeq
giving
\[
\langle k_f \vert e^{-iHt} \vert k_i \rangle =
\]
\beq
\lim_{N\to \infty}
\sum \left (\prod_n P_{X}(k_{n+1};k_n) e^{-i \sum_m \tilde{H}_2(p_{n}) \Delta t}
\right ) \delta_{k_1,k_i} \qquad k_f = k_{N+1}
\label{f25}
\eeq
which has the structure of the $N^{th}$ power of a $M \times M$ matrix.
In this discrete case the result becomes exact in the Trotter limit.

\section{wavelet basis}
\label{app3}

The construction of the wavelet basis used to construct the
discrete representation of the Hamiltonian (\ref{wav2}) is outlined below.
The starting point the solution of the linear renormalization group
equation
\beq
s(x) = \sum_{l=0}^{2L-1}h_l D T^l s(x)
\label{w3}
\eeq
%\label{wav.1}
where
\beq
Df(x):= \sqrt{2}f(2x) \qquad \mbox{and} \qquad Tf(x) := f(x-1)
%\label{wav.2}
\label{w4}
\eeq
are unitary discrete dyadic scale transformations and unit
translations.  Here $L$ is a positive integer that determines
the type of Daubechie's wavelets.
The $h_l$ are constants that depend on the choice of
$L$. Generally as $L$ increases the solutions, $s(x)$, become smoother
but the support increases.  A useful choice is $L=3$ where the
solution $s(x)$ of (\ref{w3}), called the scaling function, has
support on $[0,2L-1]=[0,5]$ and has one continuous derivative.  In
that case the coefficients $h_l$ for the Daubechies $L=3$ scaling
functions are
\[
h_0=(1+\sqrt{10}+\sqrt{5+2\sqrt{10}}\,)/16\sqrt{2}
\]
\[
h_1=(5+\sqrt{10}+3\sqrt{5+2\sqrt{10}}\,)/16\sqrt{2}
\]
\[
h_2=(10-2\sqrt{10}+2\sqrt{5+2\sqrt{10}}\,)/16\sqrt{2}
\]
\[
h_3= (10-2\sqrt{10}-2\sqrt{5+2\sqrt{10}}\,)/16\sqrt{2}
\]
\[
h_4=(5+\sqrt{10}-3\sqrt{5+2\sqrt{10}}\,)/16\sqrt{2}
\]
\beq
h_5 =(1+\sqrt{10}-\sqrt{5+2\sqrt{10}}\,)/16\sqrt{2} . 
\label{w5}
\eeq
They are determined by the requirement that the solution of (\ref{w3})
satisfies
\beq
\sum_{l=0}^{2L-1}{h_l} = \sqrt{2} 
\label{w5a}
\eeq
\beq
\int s_m(x) s_0(x) dx = \delta_{m0}
\label{w5b}
\eeq
and
\beq
x^k = \sum_{n=-\infty}^{\infty} c_n s(x-n) \qquad 0 \leq k \leq L .
\label{w5c}
\eeq
These become equations for the $h_l$ when (\ref{w3}) is used in
(\ref{w5a}),(\ref{w5b}) and (\ref{w5c}).  Equation (\ref{w5b}) means that integer translates
of $s(x)$ are orthonormal while (\ref{w5c}) means that locally
finite linear combinations of $s_n(x) = s(x-n)$ can pointwise
represent low degree polynomials.

Given the solution, $s(x)$, of (\ref{w3})
new functions are constructed from $s(x)$
by rescaling and translating
\beq
s^k_n(x) := D^k T^n (x)s(x) = \sqrt{2^k} s(2^k x-n) .
%\label{wav.6}
\label{w6}
\eeq
Since (\ref{w3}) is homogeneous in $s(x)$
the starting scale can be fixed by requiring
\beq
\int s(x) dx=1 .
\label{w7}
\eeq
The functions $s^k_n(x)$ for fixed $k$ span a subspace of the square
integrable functions on the real line with a resolution $2^{-k}L$:
\beq
{\cal S}^k := \{ f(x) \vert f(x) = \sum_{n=-\infty}^\infty c_n s^k_n(x) \qquad
\sum_{n=-\infty}^\infty  \vert c_n\vert^2 < \infty \}.
%\label{wav.5}
\label{w8}
\eeq
The renormalization group equation (\ref{w3}) implies
\beq
{\cal S}^k \subset {\cal S}^{k+1}.  
%\label{wav.11}
\label{w9}
\eeq
It follows that
\beq
{\cal S}^{k+1} = {\cal S}^{k} \oplus {\cal W}^{k}.    
%\label{wav.12}
\label{w10}
\eeq
where ${\cal W}^k$ is the orthogonal complement of
${\cal S}^k$ in ${\cal S}^{k+1}$.
An orthonormal basis for the subspace 
${\cal W}^k$ is the ``wavelet functions'':
\beq
w^k_n(x)=D^kT^n w(x)
\label{w11}
\eeq
where
\beq
w (x):= \sum_{l=0}^{2L-1} (-)^l h_{2L-1-l} D T^{l} s(x) 
%\label{wav.13}
\label{w12}
\eeq
is called the ``mother wavelet''.
This decomposition can be continued to generate a multi-resolution decomposition of $L^2(\mathbb{R})$
\beq
%\boxed{
L^2 (\mathbb{R}) = {\cal S}^{k} \oplus_{l=0}^\infty {\cal W}^{k+l}.
%=
%\oplus_{n=-\infty}^{\infty}   {\cal W}^{n}.
%\label{wav.17}
%}
\label{w13}
\eeq
This results in a multi-resolution  orthonormal basis for
$L^2 (\mathbb{R})$
\beq
\{\xi_n(x) \}_{n=-\infty}^{\infty} :=
\{ s^k_n(x)\}_{n=-\infty}^\infty \cup \{ w^m_n(x)\}_{n=-\infty,l=k}^\infty .  
%\label{wav.18}
\label{w14}
\eeq
For the choice $L=3$ the basis functions $s^k_n(x)$ and $w^k_n(x)$ have
compact support on
$[2^{-k}n,2^{-k}(n+5)]$.  All of the basis functions have one
continuous derivative so the coefficients (\ref{wav3}) are defined .  The functions $s^k_n(x)$ are like splines in
that linear combinations can be used to locally pointwise represent
degree 2 polynomials while the functions $w^l_n(x)$ are orthogonal to
the same polynomials on their support.  The Fourier transforms of the
basis functions are entire functions due to their compact support.
Orthonormal three dimensional basis functions are products of
one-dimensional basis functions.   In spite of these nice properties,
the basis functions are fractal valued (since they are related to fixed points
of a renormalization group equation) and cannot be written down in
closed form. 

\bigskip
\bigskip
\bibliographystyle{unsrt}
%\section*{References}
%\begin{thebibliography}
%\bibliography{collins.bib}
\bibliography{quant_comp.bib}

\begin{thebibliography}{10}

\bibitem{Simon}
M.~Reed and B~Simon.
\newblock {\em Methods of Modern Mathematical Physics}, volume~I.
\newblock Academic Press, San Diego, 1980.

\bibitem{Muldowney}
P.~Muldowney.
\newblock {\em A Modern Theory of Random Variation}.
\newblock Wiley, N.J., 2012.

\bibitem{Katya_1}
Ekaterina~S. Nathanson and E.T. J{\o}rgensen, Palle.
\newblock {A global solution to the Schr\o:dinger equation: From Henstock to
  Feynman}.
\newblock {\em J. Math. Phys.}, 56:092102, 2015.

\bibitem{Katya_2}
Ekaterina~S. Nathanson.
\newblock {Path integration with non-positive distributions and applications to
  the Schr\"odinger equation}.
\newblock {\em University of Iowa Thesis}, 2015.

\bibitem{Henstock}
R.~Henstock.
\newblock {\em Theory of Integration}.
\newblock Butterworths, London, 1963.

\bibitem{bartle}
R.~G. Bartle.
\newblock {\em A Modern Theory of Integration, Graduate Studies in
  Mathematics}, volume~32.
\newblock AMS, Providence, RI, 2000.

\bibitem{polyzou}
W.~N. Polyzou and Ekaterina Nathanson.
\newblock Scattering using real-time path integrals.
\newblock {\em Phys. Rev. C}, 101:064001, Jun 2020.

\bibitem{Muldowney2}
P.~Muldowney.
\newblock {Henstock on random variation}.
\newblock {\em Scientiae Mathematicae Japonicae Online}, e:657, 2007.

\bibitem{Gill}
T.~Gill and W.~Zachary.
\newblock {Bancah Spaces for the Feynman integral}.
\newblock {\em Real Analysis Exchange}, 34:1, 2008.

\bibitem{schwinger}
J.~Schwinger and B-G. Englert~ed.
\newblock {\em Quantum Mechanics}.
\newblock Springer, Berlin-Heidelberg, 2001.

\bibitem{birkhoff}
Garrett Birkhoff and John~Von Neumann.
\newblock {The Logic of Quantum Mechanics}.
\newblock {\em Annals of Mathematics}, 37(4):823--843, 1936.

\bibitem{Weyl}
H.~Weyl.
\newblock Quantenmechanik und gruppentheorie.
\newblock {\em Zeitschrift fur Physik}, 46:1, 1927.

\bibitem{brenig}
W.~Brenig and Haag R.
\newblock {General Quantum Theory of collision processes}.
\newblock {\em Fortsch.Phys.}, 7:183--242, 1959.

\bibitem{Campbell}
W.~B. Campbell, P.~Finkler, C.~E. C.~E.~Jones, and M.~N. Misheloff.
\newblock Path-integral formulation of scattering theory.
\newblock {\em Phys. Rev. D}, 12:2363, 1975.

\bibitem{Rosenfelder1}
R.~Rosenfelder.
\newblock Path integrals for potential scattering.
\newblock {\em Phys. Rev. A}, 79:012701, 2009.

\bibitem{Rosenfelder2}
J.~Carron and R.~Rosenfelder.
\newblock A new path-integral representation of the t-matrix in potential
  scattering.
\newblock {\em Phys. Letters. A}, 375:3781, 2011.

\bibitem{Brenig:1959}
W.~Brenig and R.~Haag.
\newblock {General quantum theory of collision processes}.
\newblock {\em Fort. der Physik}, 7:183, 1959.

\bibitem{Rubtsova}
O.~A. Rubtsova, V.~N. Pomerantsev, and V.~I. Kukulin.
\newblock Quantum scattering theory on the momentum lattice.
\newblock {\em Physical Review C}, 79:064602, 2009.

\bibitem{Kopp:2011vv}
P.~Kopp and W.~N. Polyzou.
\newblock {A Euclidean formulation of relativistic quantum mechanics}.
\newblock {\em Phys.Rev.}, D85:016004, 2012.

\bibitem{Haag:1958vt}
R.~Haag.
\newblock {Quantum field theories with composite particles and asymptotic
  conditions}.
\newblock {\em Phys. Rev.}, 112:669--673, 1958.

\bibitem{Ruelle:1962}
D.~Ruelle.
\newblock {On the asymptotic condition in quantum field theory}.
\newblock {\em Helv. Phys. Acta.}, 35:147, 1962.

\bibitem{jost}
R.~Jost.
\newblock {\em The General Theory of Quantized Fields}.
\newblock AMS, 1965.

\bibitem{daubechies}
I.~Daubechies.
\newblock {\em Ten Lectures on Wavelets}, volume~61.
\newblock SIAM, 1992.

\bibitem{jorgensen1}
O.~Bratteli and P.~E.~T. J{\o}rgensen.
\newblock {\em Wavelets through a looking glass, The world of the spectrum},
  volume~61.
\newblock Birkh{\"a}user, Boston, 2002.

\bibitem{jorgensen2}
P.E.T. J{\o}rgensen.
\newblock {\em Analysis and Probability, Wavelets, Signals, Fractals}, volume
  234.
\newblock Springer, NY, 2006.

\bibitem{Bulut:2013bg}
Fatih Bulut and Wayne~N. Polyzou.
\newblock {Wavelets in Field Theory}.
\newblock {\em Phys. Rev. D}, 87(11):116011, 2013.

\bibitem{Polyzou:2017wnj}
W.~N. Polyzou, Tracie Michlin, and Fatih Bulut.
\newblock {Multi-scale methods in quantum field theory}.
\newblock {\em Few Body Syst.}, 59(3):36, 2018.

\bibitem{Polyzou:2020ifj}
W.~N. Polyzou.
\newblock {Wavelet representation of light-front quantum field theory}.
\newblock {\em Phys. Rev. D}, 101(9):096004, 2020.

\bibitem{best:1994}
C.~Best and A.~Schaefer.
\newblock {Variational description of statistical field theories using
  Daubechies' wavelets}.
\newblock 1994.

\bibitem{federbush:1995}
P.~Federbush.
\newblock {New formulation and regularization of gauged %theories using a
  nonlinear wavelet expansion}.
\newblock {\em Prog. Theor. Phys.}, 94:1135, 1995.

\bibitem{1995NuPhB.436..414H}
I.~G. {Halliday} and P.~{Suranyi}.
\newblock {Simulation of field theories in wavelet representation}.
\newblock {\em Nuclear Physics B}, 436:414--427, February 1995.

\bibitem{Battle:1999}
G.~Battle.
\newblock {\em Wavelets and Renormalization, Series in Approximations and
  Decompositions, Volume 10}, volume~10.
\newblock World Scientific, 1999.

\bibitem{best:2000}
C.~Best.
\newblock {Wavelet induced renormalization group for the Landau-Ginzburg
  model}.
\newblock {\em Nucl. Phys. Proc. Suppl.}, 83:848, 2000.

\bibitem{Ismail1:2003}
A.~E. Ismail, G.~C. Rutledge, and G.~Stephanopoulos.
\newblock {Multi-Resolution analysis in statistical mechanics. I. Using
  wavelets to calculate thermodynamic properties}.
\newblock {\em J. Chem. Phys.}, 118:4414, 2003.

\bibitem{Ismail2:2003}
A.~E. Ismail, G.~C. Rutledge, and G.~Stephanopoulos.
\newblock {Multi-Resolution analysis in statistical mechanics. II. The wavelet
  transform as a basis for Monte Carlo simulations on lattices}.
\newblock {\em J. Chem. Phys.}, 118:4424, 2003.

\bibitem{altaisky:2007}
M.~V. Altaisky.
\newblock {Wavelet-Based Quantum Field Theory Symmetry, Integrability and
  Geometry: Methods and Applications}.
\newblock {\em SIGMA}, 3:105, 2007.

\bibitem{albeverio:2009}
S.~Albeverio and M.~V. Altaisky.
\newblock {A remark on gauge invariance in wavelet-based quantum field theory}.
\newblock 2009.

\bibitem{altaisky:2010}
M.~V. Altaisky.
\newblock {Quantum field theory without divergences}.
\newblock {\em Phys. Rev.}, D81:125003, 2010.

\bibitem{altaisky:2013}
M.~V. Altaisky and N.~E. Kaputkina.
\newblock {Continuous Wavelet Transform in Quantum Field Theory}.
\newblock {\em Phys. Rev.}, D88:025015, 2013.

\bibitem{altaisky:2013b}
M.~V. Altaisky and N.~E. Kaputkina.
\newblock {On the wavelet decomposition in light cone variables}.
\newblock {\em Russ. Phys. J.}, 55:1177, 2013.

\bibitem{PhysRevA.92.032315}
Gavin~K. Brennen, Peter Rohde, Barry~C. Sanders, and Sukhwinder Singh.
\newblock Multiscale quantum simulation of quantum field theory using wavelets.
\newblock {\em Phys. Rev. A}, 92:032315, Sep 2015.

\bibitem{PhysRevLett.116.140403}
Glen Evenbly and Steven~R. White.
\newblock Entanglement renormalization and wavelets.
\newblock {\em Phys. Rev. Lett.}, 116:140403, Apr 2016.

\bibitem{altaisky:2016b}
M.~V. Altaisky and N.~E. Kaputkina.
\newblock {On quantization in light-cone variables compatible with wavelet
  transform}.
\newblock {\em Int. J. Theor. Phys.}, 55:2805, 2016.

\bibitem{altaisky:2016c}
M.~V. Altaisky.
\newblock {Unifying renormalization group and the continuous wavelet
  transform}.
\newblock {\em Phys. Rev.}, D93:105043, 2016.

\bibitem{altaisky:2016}
M.~V. Altaisky.
\newblock {Wavelet view on renormalization group}.
\newblock 2016.

\bibitem{altaisky:2017}
M.~V. Altaisky.
\newblock {Wavelets and Renormalization Group in Quantum Field Theory
  Problems}.
\newblock {\em Physics of Atomic Nuclei}, 81:786, 2018.

\bibitem{Neuberger2018}
H.~Neuberger.
\newblock Wavelets and lattice field theory.
\newblock {\em EPJ Web of Conferences}, 175:11002, 2018.

\bibitem{Tomboulis1}
E.T. Tomboulis.
\newblock {Wavelet field decomposition and UV opaqueness}.
\newblock {\em J. High Energy Physics}, 77, 2021.

\bibitem{Altaisky:2021hbq}
Mikhail Altaisky, Natalia Kaputkina, and Robin Raj.
\newblock {Multiresolution quantum field theory in infinite-momentum frame}.
\newblock {\em Int. J. Theor. Phys.}, 61:46, 2022.

\bibitem{brennan3}
Mohsen Bagherimehrab, Yuval~R. Sanders, Dominic~W. Berry, Gavin~K. Brennen, and
  Barry~C. Sanders.
\newblock Nearly optimal quantum algorithm for generating the ground state of a
  free quantum field theory.
\newblock {\em PRX Quantum}, 3:020364, Jun 2022.

\bibitem{beylkin1}
G.~Beylkin.
\newblock {On the representation of operators in bases of compactly supported
  wavelets}.
\newblock {\em SIAM Journal on Numerical Analysis}, 29:1716, 1992.

\end{thebibliography}
\end{document}